\documentclass[a4paper, 12pt]{article}

\pdfoutput = 1

\usepackage{styleBuding}
\usepackage{bm}
\usepackage{color,listings}
\usepackage[usenames,dvipsnames,svgnames,table]{xcolor}
\usepackage{amsthm, amsmath}
\usepackage{amssymb}
\usepackage{mathrsfs}
\usepackage{graphicx}
\usepackage{slashed}
\usepackage{soul}
\usepackage{float}
\usepackage{multirow}
\usepackage{subfigure}
\usepackage{diagbox}
\usepackage{siunitx} 
\usepackage{url} 
\usepackage[utf8]{inputenc}
\usepackage[figuresright]{rotating}
\definecolor{back}{HTML}{F8F8F8}
\usepackage{epstopdf}
\usepackage{lipsum}

\usepackage{booktabs}  
\usepackage{array}  
\usepackage{fancyhdr} 

\newcommand\blfootnote[1]{%
	\begingroup
	\renewcommand\thefootnote{}\footnote{#1}%
	\addtocounter{footnote}{-1}%
	\endgroup
}

\newcommand{\rom}[1]{\uppercase\expandafter{\romannumeral #1\relax}}

\allowdisplaybreaks
\DeclareUnicodeCharacter{2212}{-}

\title{Higgsino Dark Matter Interpretation of the LZ High-Recoil Event in the GNMSSM with TeV-Scale Gauginos}

\author{Subhadip Bisal, Junjie Cao$^*$, and Fei Li$^*$\blfootnote{*Corresponding author.}}

\affiliation{School of Physics, Zhengzhou University, Zhengzhou 450000, China}

\emailAdd{subhadipbisal6@gmail.com}
\emailAdd{junjiec@alumni.itp.ac.cn}
\emailAdd{hnufeili@163.com}

\abstract{The nuclear recoil event at approximately $248~{\rm keV}$ reported
by the LUX--ZEPLIN  (LZ) collaboration motivates an investigation of
endothermic dark matter (DM) scattering. We study this interpretation
within the General Next-to-Minimal Supersymmetric Standard Model (GNMSSM), 
with Higgsino-dominated neutralino DM undergoing
the $Z$-mediated transition
$\widetilde{\chi}_1^0N\to\widetilde{\chi}_2^0N$.
In the conventional thermal, nearly pure-Higgsino limit of the
Minimal Supersymmetric Standard Model, the observed relic abundance
selects a mass near $1.1~{\rm TeV}$, while a neutralino splitting of
a few hundred keV typically requires electroweak Gaugino masses
of order $10^7~{\rm GeV}$ in the absence of cancellations.
In the GNMSSM, Higgsino--Singlino mixing introduces an additional
contribution to the splitting that can cancel against the
Gaugino-induced contribution, allowing a sub-MeV splitting with
multi-TeV Gauginos. This mixing also modifies the inelastic
scattering coupling and the thermal annihilation rates, while
coannihilation with nearby Sleptons provides additional freedom
in obtaining the observed relic abundance. We present six
benchmark points of our previous work with physical DM masses of approximately
$0.6$--$1.1~{\rm TeV}$, neutralino splittings of
$330$--$350~{\rm keV}$, and Gaugino mass parameters of magnitude
$2$--$5~{\rm TeV}$. These points reproduce the observed relic
abundance and satisfy the direct-detection, Higgs, flavor, and
collider constraints considered in that work. Within the adopted
Standard Halo Model and extended likelihood analysis, all six
points yield $\Delta\chi^2<1$ relative to the reference best fit 
for the new experimental results.
Our results illustrate how the GNMSSM can accommodate the LZ
high-recoil event without requiring an ultraheavy Gaugino sector.
A quantitative assessment of solar-capture and neutrino-telescope
constraints remains necessary to establish the viability of this
interpretation.  

}

\begin{document}
    \maketitle
    \flushbottom


\section{Introduction}
\label{sec:introduction}

The identification of the particle nature of dark matter (DM) remains one of the central problems in particle physics and cosmology. 
Recently, the LUX--ZEPLIN (LZ) collaboration reported the observation of a
single candidate DM event with a nuclear recoil energy of
$E_R = 248 \pm 23\,(\text{stat.}) \pm 23\,(\text{sys.})~\text{keV}$, recorded on
June 16, 2023, in a search that extended the nuclear recoil energy window up to
$269.9~\text{keV}$ with an exposure of $2.84~\text{tonne-years}$~\cite{LZ:2026axp}. 
The event occurred in a region characterized by an exceptionally low expected 
background (approximately $0.01$ events), and no known instrumental or radiogenic process has been identified 
that could plausibly explain it.
A profile likelihood ratio analysis quantifies the
tension with the background-only hypothesis at a global significance of
$2.6\sigma$ after accounting for look-elsewhere effects, with a maximum local
significance of $3.4\sigma$ among the signal models tested. 

Although far from
discovery level, the event is intriguing precisely because of its kinematics:
conventional spin-independent (SI) elastic scattering of a weakly interacting
massive particle (WIMP) produces a recoil spectrum that falls steeply with
energy, so that a single event at $\sim 250~{\rm keV}$ would have to be
accompanied by thousands of events at low recoil energies, in stark conflict
with the null result of the standard LZ search. The observation therefore
points toward a scattering mechanism whose spectrum is intrinsically peaked at
high energies. The most economical such mechanism is \emph{inelastic} DM
scattering~\cite{Bramante:2016rdh}, in which the DM particle $\chi_1$ up-scatters
into a slightly heavier partner $\chi_2$ with mass splitting
$\delta \equiv m_{\chi_2} - m_{\chi_1}$. Kinematics then requires an incoming
velocity $v \gtrsim v_{\rm min} \simeq \sqrt{2\delta/\mu_A}$, where $\mu_A$ is
the DM--nucleus reduced mass, and the typical recoil energy is
$E_R \simeq (\mu_A/m_A)\,\delta$, essentially independent of the DM mass.
The LZ event thus translates directly into a preferred splitting of a few
hundred keV; for a xenon target, the precise value depends on the high-velocity
tail of the local DM velocity distribution, ranging from
$\delta \simeq 350~{\rm keV}$ for the Standard Halo Model (SHM) to
$\delta \simeq 500~{\rm keV}$ when a high-velocity component associated with
the Large Magellanic Cloud (LMC) is included~\cite{Fan:2026kxx,Freese:2026sga}. 

The observed excess has attracted significant theoretical interest, 
prompting investigations across multiple theoretical 
frameworks~\cite{Fan:2026kxx,Du:2026guj,Freese:2026sga,Wu:2026nhi,
Pospelov:2026ewn,Rodd:2026tyn,Su:2026rwz,Lou:2026idn,Yamashita:2026ump,Visinelli:2026kgt,
DiMauro:2026ldr,Unwin:2026rdp,Jeesun:2026vzo,McCabe:2026crm,Smirnov:2026aqk,
Chattopadhyay:2026ryw,Dent:2026bji,deLima:2026shq,Gu:2026vto,Das:2026uyy,
Yin:2026jnn,Bose:2026ndd,Wang:2026ytg,Yang:2026wpb,Liang:2026coz,Du:2026lpa,
Khan:2026nwp,Cheung:2026byg,Yuan:2026djt,Zhu:2026dag,Langhoff:2026ujr,Lee:2026jxl,Frolovsky:2026tvq,DiMauro:2026dqp,Nguyen:2026lui, Borah:2026zwf, Bandyopadhyay:2026gjw}.  Among the candidates
capable of realizing the high-energy recoil, the Higgsino stands
out as arguably the best motivated~\cite{Fan:2026kxx,Du:2026guj,Wu:2026nhi,Freese:2026sga,Langhoff:2026ujr,Frolovsky:2026tvq,
Griest:2000kj, Kane:1996dd, Drees:1996pk, Masip:2005fv, Wang:2005kf, Hall:2009aj,Cao:2016nix, 
Kowalska:2018toh, Delgado:2020url, Yue:2025dqe, Bisal:2023fgb, Bisal:2024ezn, 
Bisal:2026hpm,Fox:2014moa, Chattopadhyay:2005mv, Chakraborti:2014fha, 
Shafi:2023sqa, Mummidi:2018myd,Nagata:2014wma, Co:2021ion, Martin:2024pxx, 
Martin:2024ytt,Chun:2016cnm,Yue:2025wnm}. 
In supersymmetric (SUSY) theories, Higgsinos are the fermionic partners 
of the Higgs doublets. In the minimal construction, two Higgs doublets 
are required to generate both up-type and down-type fermion masses 
through a holomorphic superpotential after the electroweak symmetry 
breaking (EWSB) and to cancel the gauge anomalies associated with their 
fermionic partners~\cite{Haber:1984rc,Gunion:1984yn}. Higgsinos therefore 
arise independently of the interpretation of the LZ event.
Their suitability for inelastic scattering follows from the structure of the neutralino
mass matrix. In the limit $v_{\rm EW}\to 0$, the tree-level neutralino mass terms
respect a global $U(1)$ Higgsino number under which $\widetilde H_u^0$ and
$\widetilde H_d^0$ carry opposite charges, so that the two neutral Higgsinos combine
into a single Dirac fermion, or equivalently into two exactly degenerate Majorana
states. After EWSB, interactions that violate this $U(1)$ split the Dirac fermion
into two Majorana mass eigenstates $\widetilde\chi^0_1$ and $\widetilde\chi^0_2$ with
$\delta\equiv|m_{\widetilde\chi^0_2}|-|m_{\widetilde\chi^0_1}|$. The smallness of
$\delta$ relative to the Higgsino mass is therefore not accidental: it reflects
the hierarchy $v_{\rm EW}\ll\Lambda_{\rm SUSY}$, and, in the absence of
accidental cancellations, $\delta$ serves as an order parameter for the breaking
of Higgsino number. The same structure controls the couplings to the $Z$ boson.
Since the vector current of a Majorana fermion vanishes identically, the
$Z$ boson couples to the Higgsino-like states predominantly through the
off-diagonal combination $\widetilde\chi^0_1\widetilde\chi^0_2$, which induces the
endothermic process $\widetilde\chi^0_1 N\to\widetilde\chi^0_2 N$; for a splitting of a
few hundred keV, this process probes the high-velocity tail of the Galactic DM
distribution and shifts the recoil spectrum toward high energies~\cite{Fan:2026kxx,Freese:2026sga}.
The diagonal axial coupling to $\widetilde\chi^0_1\widetilde\chi^0_1$, proportional to
the disparity between the $\widetilde{H}_u^0$ and $\widetilde{H}_d^0$ components in $\widetilde\chi^0_1$, 
and the SM-like Higgs coupling, which requires Higgsino--Gaugino mixing, are 
both suppressed in the pure-Higgsino limit.
Consequently, the spin-dependent (SD) and SI elastic scattering of Higgsino-like
DM off nucleons are strongly suppressed, and such candidates readily satisfy the
latest LZ exclusion limits~\cite{Martin:2024pxx, Martin:2024ytt, Yue:2025dqe, Yue:2025wnm}.
Given these features, several recent
studies have examined whether the LZ event can be interpreted as the inelastic
scattering of an approximately $1.1$~TeV Higgsino~\cite{Fan:2026kxx,Freese:2026sga,Du:2026guj,Wu:2026nhi,Langhoff:2026ujr,Frolovsky:2026tvq}, 
the mass preferred by conventional thermal freeze-out for the measured relic abundance
$\Omega h^2=0.12$~\cite{WMAP:2012nax, Planck:2018vyg}.

Within the Minimal Supersymmetric Standard Model (MSSM), however, this
interpretation is highly restrictive. The reason is structural: once the DM is
required to be a nearly pure Higgsino, its mass, its scattering couplings, and
the splitting $\delta$ are all fixed by a small number of parameters, and the
splitting in particular is controlled by nothing but the electroweak Gaugino
masses. Three rigidities follow.
\begin{itemize}
\item The splitting is tied to the Gaugino sector. In the MSSM, the sole
source of Higgsino-number breaking in the neutralino sector is the
Gaugino--Higgsino--Higgs interaction, which mixes the neutral Higgsinos with the
Bino and Wino after EWSB. At leading order in the expansion parameter 
$v_{\rm EW}/M_{1,2}$, the MSSM contribution is approximated by  
\begin{equation}
\delta_{\rm MSSM} \approx m_Z^2 \left| \frac{\sin^2\theta_W}{M_1} + \frac{\cos^2\theta_W}{M_2} \right|,
\label{eq:deltaMSSM}
\end{equation}  
with higher-order corrections suppressed by additional powers of 
$v_{\rm EW}/M_{1,2}$. Here, $M_1$ and $M_2$ denote the Bino and Wino mass parameters, 
respectively, and $\theta_W$ is the weak mixing angle.
For same-sign Gaugino masses of comparable magnitudes,  a
splitting of $\mathcal{O}(100)$~keV requires $|M_{1,2}|\sim 10^6$--$10^7$~GeV,
far above the electroweak scale~\cite{Nagata:2014wma,Chun:2016cnm,Fan:2026kxx}. 
Radiative stability of a weak-scale Higgsino mass in the presence of 
such heavy Gauginos further correlates the heavy-Higgs sector with this intermediate 
scale~\cite{Du:2026guj}. The resulting spectrum is rather special: the 
Higgsinos remain near the TeV scale, while the electroweak Gauginos, and in 
representative  ultraviolet (UV) realizations also 
the heavy Higgs states, are many orders of magnitude heavier. In this case, 
the model-building burden shifts to explaining why
$\mu^2 \ll b\mu \sim M_{\rm SUSY}^2$~\cite{Fan:2026kxx}.
The alternative is to keep
$|M_{1,2}|$ at a few TeV and arrange a cancellation between the Bino and Wino
terms in Eq.~\eqref{eq:deltaMSSM}, which at the Higgsino scale requires
$M_1/M_2\simeq-\tan^2\theta_W$ to per-mille accuracy. Such a relation does not
follow from universal boundary conditions; it must be imposed through a
non-universal input at the mediation scale and preserved under
renormalization-group evolution down to the Higgsino scale~\cite{Du:2026guj}. 
Either way, a strongly hierarchical or finely arranged Gaugino sector is an
unavoidable ingredient rather than an incidental feature. This is problematic
for UV model building because $M_1$ and $M_2$ have broad implications for the
theory: they enter the renormalization-group evolution of the entire soft
sector and are frequently related to one another and to the Gluino mass by the
mediation mechanism. Fixing them by a single sub-MeV observable therefore
propagates into essentially every other sector of the model, which usually 
distorts severely the underlying mediation mechanism and induces large, 
unwanted quantum corrections across the entire sparticle spectrum. It is this
structural entanglement, rather than the numerical size of $M_{1,2}$ alone,
that constitutes the central obstacle to the MSSM interpretation.

\item The DM mass is fixed by cosmology. In the absence of additional states 
or interactions that materially alter freeze-out, the annihilation cross
section of a pure Higgsino is determined entirely by electroweak gauge
couplings. Consequently, thermal relic abundance pins the Higgsino mass to
$\mu\simeq1.1$~TeV~\cite{Delgado:2020url, Fox:2014moa, Chattopadhyay:2005mv, Chakraborti:2014fha, Shafi:2023sqa, Mummidi:2018myd}. 

\item The scattering rates are fixed by the same parameters. The
off-diagonal $Z\widetilde\chi^0_1\widetilde\chi^0_2$ coupling approaches gauge strength
in the pure-Higgsino limit. As a result, once the DM mass and local abundance are
specified, the normalization of the inelastic scattering rate is essentially
fixed and $\delta$ remains the only parameter controlling the LZ signal. The
elastic SI and SD cross sections are likewise not adjustable at will: they are
governed by the same Higgsino--Gaugino mixing that generates $\delta$, so that
the direct-detection rates and the splitting are locked together through
$M_1$ and $M_2$.
\end{itemize}

The MSSM interpretation is also subject to important complementary constraints.
Higgsinos falling into the solar gravitational potential acquire velocities
substantially larger than those available to halo particles in terrestrial
detectors, e.g.\ $v\gtrsim1300$~km/s in the solar core, and may therefore
up-scatter on heavy solar elements even when the corresponding process is
kinematically suppressed on Earth. Their subsequent capture and annihilation
can produce energetic neutrinos. Under the assumptions of efficient
post-capture thermalization, capture--annihilation equilibrium, and standard
solar and halo models, a recent analysis of IceCube data obtained the
requirement $\delta\gtrsim566$~keV for a $1.08$~TeV Higgsino
~\cite{Pospelov:2026ewn,Bose:2026ndd,DiMauro:2026dqp,Nguyen:2026lui}, in
pronounced tension with the splitting preferred by the LZ event. 
The numerical value of this bound nevertheless depends on the treatment 
of solar capture and thermalization, on the abundances of the relevant 
heavy elements, and on the assumed annihilation channels. It should 
therefore be regarded as a powerful model-dependent constraint rather 
than a model-independent exclusion of every Higgsino realization. 
In addition, the absence of events in the higher-energy 
LZ sideband ($800 < S1_c < 1700~\text{phd}$) has been argued to disfavor parts of the 
thermal-Higgsino parameter space~\cite{Rodd:2026tyn}, and indirect searches using gamma 
rays and collider probes provide further complementary tests~\cite{Wu:2026nhi}. 

Taken together, these considerations highlight that the primary obstacles to interpreting 
the LZ excess do not stem from the Higgsino hypothesis itself, but rather from the 
rigid correlations inherent to the MSSM: a single DM mass, a fixed cross section,
and a splitting controlled by nothing but a pair of very heavy or precisely
correlated Gaugino masses. It is therefore natural to ask whether Higgsino 
DM in an extended SUSY framework can accommodate the LZ event while softening these
rigidities. What is required is not merely additional parameters, but an
independent source of Higgsino-number breaking that can generate the required
splitting without pushing the Gauginos to an intermediate scale or imposing a
special relation between them, together with enough freedom in the DM mass to
depart from the $1.1$~TeV thermal prediction.

The General Next-to-Minimal Supersymmetric Standard Model (GNMSSM) provides
such a framework. In addition to the two Higgs doublet superfields, it contains
a gauge-singlet superfield $\widehat S$, whose fermionic component, the Singlino,
enlarges the neutralino sector~\cite{Ellwanger:2009dp,Hollik:2018yek,Hollik:2020plc,
Cao:2021ljw,Cao:2024axg,Meng:2024lmi}. The superpotential coupling 
$\lambda\widehat S\widehat H_u\cdot\widehat H_d$, in conjunction 
with the general singlet-sector mass parameters, introduces a second source of
Higgsino-number breaking, which has non-trivial implications on each of the rigidities 
after EWSB. 
Crucially, the resulting Higgsino–Singlino mixing in the neutralino mass 
matrix yields a potentially substantial contribution to the neutral-Higgsino mass splitting. 
The magnitude and sign of this contribution remain independent 
of the Gaugino mass parameters, and are governed solely by $\lambda$ and 
the singlet-sector parameter denoted by $d$ in this work.  
This singlet-induced contribution interferes with the Gaugino-induced term in
Eq.~\eqref{eq:deltaMSSM}, thereby transforming the phenomenological constraint 
$\delta \simeq 350~{\rm keV}$ from a simple relation between $M_1$ and $M_2$ 
into a matching condition linking the Gaugino and singlet sectors. As we 
demonstrate below, the observed splitting can then be obtained for 
multi-TeV Gauginos of either relative sign without requiring the Bino 
and Wino contributions to cancel each other. 
We emphasize that this does not remove the need for a precise cancellation, since
a sub-MeV splitting is still generated from TeV-scale inputs; what changes is
that the sensitivity is carried by $\lambda$ and $d$, parameters directly
constrained by the DM observables themselves, rather than by parameters
entangled with the UV structure of the soft sector. In this sense the GNMSSM
relaxes a restrictive condition on the Gaugino sector and increases the
flexibility of UV model building, although it does not by itself eliminate
fine tuning.

Higgsino--Singlino mixing also relaxes the remaining rigid predictions of the
MSSM scenario. If the DM state contains a Singlino fraction $\sin\theta$, its
off-diagonal coupling to the $Z$ boson, and hence the inelastic scattering
cross section, are reduced relative to the pure-Higgsino values by factors of
$\cos\theta$ and $\cos^2\theta$, respectively~\cite{Yue:2025dqe}. 
In the elastic SI
channel, the several Higgs-mediated contributions, now including the
singlet-like states, can interfere destructively, suppressing
$\sigma^{\rm SI}_N$ well below the MSSM value at the same DM
mass~\cite{Yue:2025dqe}. Both effects loosen the one-to-one correspondence between the
observed event rate and $\delta$ that characterizes the MSSM. 
Moreover, the same mixing modifies the annihilation and coannihilation processes that determine
the relic abundance and opens channels involving singlet-like Higgs bosons.
Higgsino-dominated neutralino DM in the GNMSSM has been shown to reproduce the
measured relic abundance through thermal freeze-out for masses down to roughly
$700$~GeV while satisfying current direct-detection constraints~\cite{Yue:2025dqe}, and
if coannihilation with nearly degenerate Sleptons is operative, viable masses
may extend to approximately $300$~GeV~\cite{Chakraborti:2017dpu,Yue:2025wnm}. 
The DM mass is thus no longer fixed at $1.1$~TeV by cosmology. 
This freedom is phenomenologically relevant for the complementary 
constraints discussed above. Lighter Higgsinos have been
noted to reduce the tension with the LZ high-energy sideband~\cite{Rodd:2026tyn}.
Regarding solar capture, the IceCube bound derived for a $1.08$~TeV pure
Higgsino cannot be transplanted directly to a lighter, mixed state: the
inelastic coupling is reduced by $\cos^2\theta$, the incident number density
scales as $\rho_0/m_{\widetilde\chi^0_1}$, and the kinematic threshold for
up-scattering in the Sun shifts with the DM mass. The constraint must therefore
be re-evaluated for each parameter point, a question to which we return in
Sec.~\ref{Solar-neutrino}.

Motivated by these observations, in this work we investigate whether
Higgsino-dominated neutralino DM in the GNMSSM can account for the LZ
high-energy recoil event with a multi-TeV Gaugino sector. 
We analyze the origin
of the small neutralino splitting and its parameter
sensitivity, and examine how Higgsino--Singlino mixing
and Slepton coannihilation affect the relic abundance
and inelastic scattering rate. Six representative
benchmark points from our previous study illustrate
this possibility for both same-sign and opposite-sign
Gaugino mass parameters. Their agreement with the LZ
event is evaluated within the adopted halo model and
likelihood. Finally, we scrutinize the complementary solar 
neutrino constraints, identifying the concrete pathways 
through which the GNMSSM can reconcile the signal 
with global limits.

The remainder of this paper is organized as follows. Sec.~\ref{sec:theory} describes the
kinematics of endothermic scattering relevant to the LZ event, the calculation
of the event rate, and the likelihood used to compare theoretical predictions
with the data. Sec.~\ref{sec:gnmssm} introduces the GNMSSM, analyzes its neutralino sector
with emphasis on Higgsino--Singlino mixing and the resulting mass splitting,
assesses the associated fine tuning, and summarizes the inelastic scattering
cross section and the relic abundance. Sec.~\ref{sec:lz_fit} presents the numerical analysis,
in which six benchmark points from our previous work are confronted with the LZ
event and their properties are compared with the MSSM expectations. Sec.~\ref{Solar-neutrino}
discusses the solar neutrino constraints and possible suppression mechanisms
within the GNMSSM. We summarize our conclusions in Sec.~\ref{sec:conclusion}.

\section{\label{theory-section} The LZ High-Recoil Event: kinematics and Likelihood} 
\label{sec:theory}
In this section, we describe the kinematics of endothermic DM scattering
and the procedure used to calculate the corresponding event spectrum in the LZ detector. We then define the likelihood employed to compare the
theoretical predictions with the observed high-recoil event.

\subsection{Kinematics of Inelastic Scattering}
\label{sec:kinematics}

We interpret the LZ event as a candidate DM--nucleus scattering and examine the corresponding kinematic constraints. The canonical SI elastic scattering scenario, absent form-factor suppression or momentum-dependent couplings, predicts a recoil spectrum concentrated at low energies. This picture is in
acute tension with a signal at $E_R = 248$~keV: reproducing one such event would
require thousands of accompanying events at low recoil energy, in stark conflict
with the null result of the standard LZ search.

The tension is resolved if the scattering is endothermic,
$\widetilde{\chi}_1^0 + N \to \widetilde{\chi}_2^0 + N$, with a mass splitting
\begin{equation}
  \delta \;\equiv\; |m_{\widetilde{\chi}_2^0}| - |m_{\widetilde{\chi}_1^0}| \;>\; 0 .
  \label{eq:delta_def}
\end{equation}

In this paradigm, the threshold velocity for producing a recoil of energy $E_R$ is set by kinematic conservation laws and the mass-splitting parameter. The minimum incoming speed $v_{\rm min}$ required for such a collision can be expressed as:

\begin{align}
	v_{\rm min} = \sqrt{\frac{1}{2m_N E_R}} \left(\frac{m_N E_R}{\mu_A} + \delta\right)~,
    \label{eq:vmin}
\end{align}
where $m_N\simeq 122$~GeV is the target nucleus mass (${}^{131}\text{Xe}$ in the LZ detector), $\mu_A = m_{\widetilde{\chi}_1^0} m_N / (m_{\widetilde{\chi}_1^0} + m_N)$ is the
DM--nucleus reduced mass, and $\delta$ parameterizes the mass splitting between Higgsino mass eigenstates that governs the inelastic scattering kinematics.

\begin{figure}
	\centering
	\includegraphics[width=0.55\linewidth]{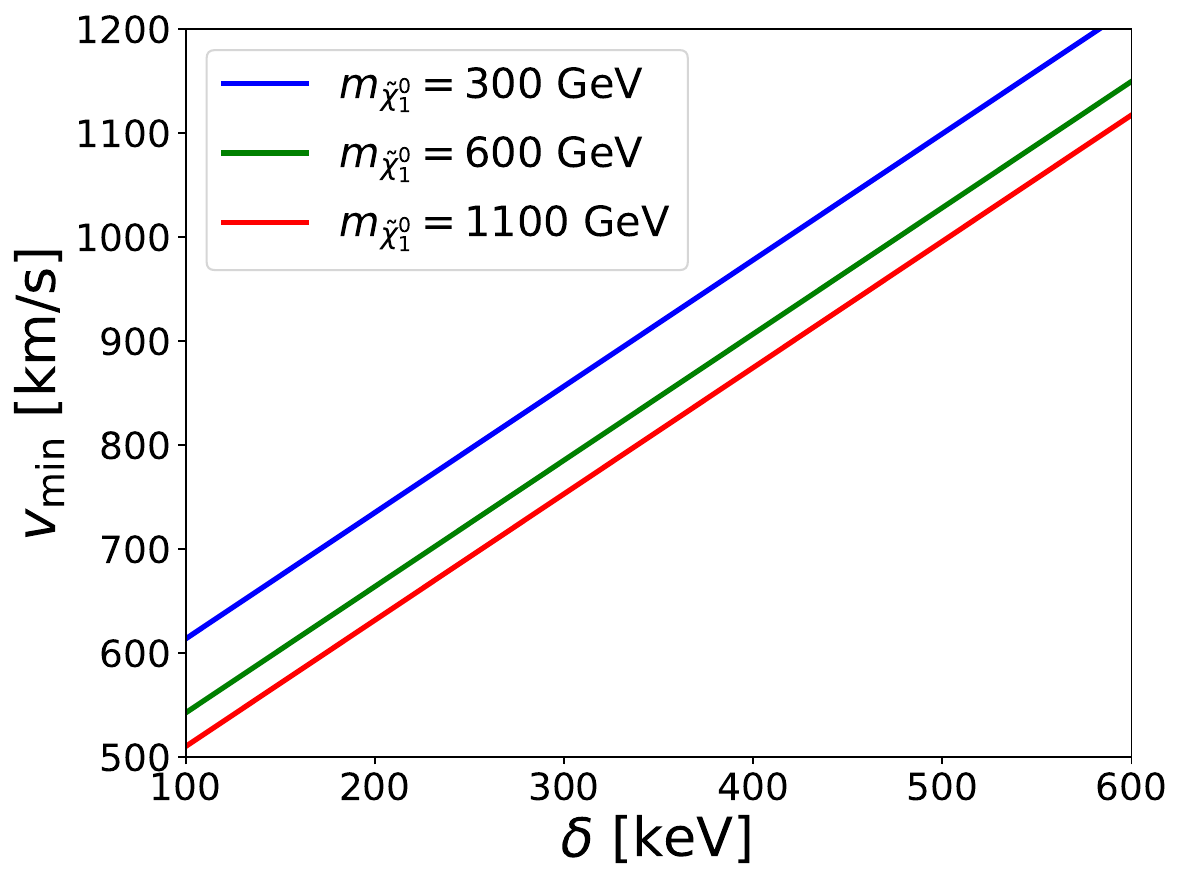}
	\caption{Dependence of minimum velocity $v_{\rm min}$ on the mass splitting $\delta$ for DM masses $m_{\widetilde{\chi}_1^0}=300$~GeV, $600$~GeV, and $1100$~GeV evaluated at $E_R = 250$~keV for the LZ detector.}
	\label{fig:vminvsdelta}
\end{figure}

Eq.~\eqref{eq:vmin} reveals that a positive mass splitting selects the high-velocity tail 
of the local DM distribution, allowing the recoil spectrum to concentrate at energies 
significantly higher than those typical of elastic weakly interacting massive particle 
scattering. A splitting of a few hundred keV thus produces an isolated recoil near 250~keV 
without generating a large event rate in the conventional low-energy search region. 
Figure~\ref{fig:vminvsdelta} demonstrates that this effect becomes more pronounced for lighter 
DM candidates: mass splittings of $\delta\gtrsim 250$~keV require minimum velocities of $v_{\rm min}\gtrsim 700$~km/s, indicating that the DM responsible for the observed scattering 
event must originate from the high-velocity tail of the local halo distribution.

DM in the galactic halo exhibits a broad spectrum of velocities, conventionally modeled through modifications to the Maxwell-Boltzmann distribution. The Standard Halo Model (SHM) represents the minimal and most widely adopted framework, characterized by a truncated Maxwell-Boltzmann velocity distribution in the galactic rest frame:
\begin{align*}
	f_{\rm SHM}(\vec{v}) \propto \exp\left(-\frac{|\vec{v}|^2}{v_0^2}\right) \Theta\left(v_{\rm esc} - |\vec{v}|\right)~,
\end{align*}
where $v_0=220$~km/s denotes the velocity dispersion, $v_{\rm esc}=540$~km/s is the Milky Way escape velocity, and $\Theta$ is the Heaviside step function.

\subsection{Event Rate}
\label{sec:eventrate}

The differential event rate per unit detector mass is given by the velocity-averaged flux:~\cite{Tucker-Smith:2001myb, Tucker-Smith:2004mxa}:
\begin{align}
	\frac{dR}{dE_R} = N_T \frac{\rho_{\widetilde{\chi}_1^0}}{m_{\widetilde{\chi}_1^0}} \int_{v>v_{\rm min}}^{} d^3v\, v\, f\left(\vec{v} + \vec{v}_e\right) \frac{d\sigma}{dE_R}~,
\end{align}
where $N_T$ is the number density of target nuclei,  $\rho_{\widetilde{\chi}_1^0}=0.4~\text{GeV}/\text{cm}^3$ is the local DM mass density (assuming Higgsino comprises the entire DM), and $m_{\widetilde{\chi}_1^0}$ is the mass of Higgsino-like DM. 

For the coherent $Z$-mediated inelastic scattering, the nuclear differential cross section is written as
\begin{align}
	\frac{d\sigma}{dE_R} = \frac{m_N}{2 v^2} \frac{\sigma_n}{\mu_A^2} \left[\frac{\left\{Z f_p + (A-Z) f_n\right\}^2}{f_n^2}\right] \mathcal{F}^2(E_R)~,
    \label{eq:dsigmadER}
\end{align}
where $\sigma_n$ is the zero-momentum-transfer reference cross section per neutron, which characterizes the strength of the microscopic inelastic interaction and $A$ and $Z$ are the nuclear atomic mass and number, respectively. Here, $f_p$ and $f_n$ denote the effective vector couplings of the DM to the proton and neutron, respectively. 

Crucially, $\sigma_n$ in Eq.~(\ref{eq:dsigmadER}) is the reference inelastic scattering cross section per neutron evaluated in the zero-momentum-transfer limit ($q^2 \to 0$):
\begin{equation}
    \sigma_n \equiv \frac{\mu_n^2}{\pi} |f_n|^2,
    \label{eq:sigma_n_definition}
\end{equation}
where $\mu_n = m_{\widetilde{\chi}_1^0} m_n / (m_{\widetilde{\chi}_1^0} + m_n) \simeq m_n$ denotes the DM--neutron reduced mass. 

In electroweak scenarios where the transition $\widetilde{\chi}_1^0 \to \widetilde{\chi}_2^0$ is mediated by the Standard Model $Z$ boson, the neutral current couples to the valence quark vector currents $v_q = T_3^q - 2 Q_q \sin^2\theta_W$, yielding:
\begin{equation}
    \frac{f_p}{f_n} = -\left(1 - 4\sin^2\theta_W\right) \simeq -0.08,
    \label{eq:fp_fn_ratio}
\end{equation}
with the weak mixing angle $\sin^2\theta_W \simeq 0.231$. Because of this strong accidental cancellation for the proton, the coherent nuclear scattering amplitude is predominantly dictated by the neutron content of the nucleus, $[(A-Z) - 0.08 Z]^2$. The absolute normalization contained in $f_n$ is absorbed into $\sigma_n$; its GNMSSM expression will be derived in Sec.~\ref{sec:inelastic_scattering}.

%
The nuclear form factor $\mathcal{F}(E_R)$ is taken in the Helm parametrization~\cite{Helm:1956zz, Engel:1991wq}:
\begin{align}
	\mathcal{F}^2(E_R) = \left(\frac{3j_1(qr_0)}{qr_0}\right)^2 \exp[-s^2 q^2], 
\end{align}
with $q=\sqrt{2m_N E_R}$, $r_0=\sqrt{r^2 - 5s^2}$, $r=1.2 A^{1/3}$~fm, and $s= 1$~fm.


\subsection{Likelihood Construction}
\label{sec:likelihood}

To assess whether the observed LZ event is consistent with the Higgsino-like DM explanation in the GNMSSM, we employ the extended maximum likelihood formalism, expressed as~\cite{Barlow:1990vc, Fan:2026kxx}:
\begin{align}
	\mathcal{L}\left(\boldsymbol{p}\right) = \exp\left[-N_{\rm ev}\left(\boldsymbol{p}\right)\right] \prod_{i=1}^{N_0} \frac{dN\left(\boldsymbol{p}\right)}{dE_R^\prime}\bigg|_{E_R^\prime=E_i}~,
\end{align}
where $N_{\rm ev}$ is the total number of expected events integrated across the detector's recoil energy range:
\begin{align}
	N_{\rm ev} = \int_{E_R^{\rm min}}^{E_R^{\rm max}} \frac{dN\left(\boldsymbol{p}\right)}{dE_R^\prime} dE_R^\prime~,
\end{align}
where
\begin{align}
    E^\prime_{\rm min} = 5.4~\text{keV}, \quad\quad E^\prime_{\rm max} = 269.9~\text{keV}~,
\end{align}
and $\boldsymbol{p}$ denotes the model parameters.
The differential event rate is $\frac{dN}{dE_R} = \frac{dR}{dE_R}\times \mathcal{E}$, where $\mathcal{E}=2.84$ tonne-year is the LZ exposure. In our case, $N_0=1$ corresponds to the single observed event at $E_R=248\pm 23 (\text{stat.}) \pm 23 (\text{sys.})$~keV.

We characterize the relative quality of a parameter point by
\begin{align}
    \Delta \chi^2 (\boldsymbol{p}) = -2 \log \left[\frac{\mathcal{L}(\boldsymbol{p})}{\mathcal{L}_{\rm max}}\right]~.
\end{align}

This quantity measures the agreement relative to the best-fitting point within the
likelihood defined above. We stress that
$\Delta \chi^2$ measures proximity to that best fit within the simplified
likelihood defined above, and does not by itself constitute a full statistical
test including detector response, acceptance and background modelling.

\section{Higgsino DM in the GNMSSM}
\label{sec:gnmssm}

Having identified $\delta \sim \mathcal{O}(100)$~keV together with a coherent,
unsuppressed inelastic coupling as the two requirements imposed by the LZ event,
we now show how both arise naturally in the GNMSSM. We introduce the model and our
parameter conventions in Sec.~\ref{sec:model}, analyze the neutralino sector and
the origin of the fine splitting in Sec.~\ref{sec:neutralino_sector}, derive the
inelastic cross section in Sec.~\ref{sec:inelastic_scattering}, and close with a brief
discussion of the relic abundance in Sec.~\ref{subsec:relic_density}.

\subsection{The Model and Parameter Conventions}
\label{sec:model}

The GNMSSM extends the MSSM by a gauge-singlet superfield $\widehat{S}$, carrying
neither baryon nor lepton number. The Higgs sector thus consists of $\widehat{S}$ and
the two $SU(2)_L$ doublets $\widehat{H}_u = (\widehat{H}_u^+, \widehat{H}_u^0)$ and
$\widehat{H}_d = (\widehat{H}_d^0, \widehat{H}_d^-)$, with superpotential~\cite{Ellwanger:2009dp}
\begin{equation}
  W_{\rm GNMSSM} = W_{\rm Yukawa} + \lambda \widehat{S} \widehat{H}_u \!\cdot\! \widehat{H}_d
  + \frac{\kappa}{3} \widehat{S}^3 + \mu \widehat{H}_u \!\cdot\! \widehat{H}_d
  + \frac{1}{2} \mu' \widehat{S}^2 + \xi \widehat{S},
  \label{eq:superpotential}
\end{equation}
where $W_{\rm Yukawa}$ contains the MSSM quark and lepton Yukawa couplings. The
terms proportional to $\mu$, $\mu'$ and $\xi$ address the tadpole
problem~\cite{Ellwanger:1983mg, Ellwanger:2009dp} and the cosmological domain-wall
issue~\cite{Abel:1996cr, Kolda:1998rm, Panagiotakopoulos:1998yw}; since a shift of $\widehat{S}$ together with a redefinition of
parameters removes one of them~\cite{Ross:2011xv}, we set $\xi = 0$ without loss of
generality. Naturalness arguments based on the breaking of discrete $R$-symmetries
$\mathbb{Z}_4^R$ or $\mathbb{Z}_8^R$ at high scales place $\mu$ and $\mu'$ in the
few-hundred-GeV rang~\cite{Abel:1996cr, Lee:2010gv, Lee:2011dya, Ross:2011xv, Ross:2012nr}. In contrast to the MSSM, the GNMSSM
accommodates both a bare $\mu$-term and a dynamically generated contribution, so
that the effective Higgsino mass is
\begin{equation}
  \mu_{\rm tot} = \mu + \lambda v_s/\sqrt{2},
  \qquad \langle S \rangle \equiv v_s/\sqrt{2}.
  \label{eq:mutot}
\end{equation}
The corresponding soft-breaking Lagrangian in the Higgs sector reads
\begin{eqnarray}
    -\mathcal{L}_{soft} = &\Bigg[\lambda A_{\lambda} S H_u \cdot H_d + \frac{1}{3} \kappa A_{\kappa} S^3+ m_3^2 H_u\cdot H_d + \frac{1}{2} {m_S^{\prime}}^2 S^2 + \xi^\prime S + h.c.\Bigg]  \nonumber \\
   & + m^2_{H_u}|H_u|^2 + m^2_{H_d}|H_d|^2 + m^2_{S}|S|^2 , \label{Soft-terms}
\end{eqnarray}

Throughout this work, and in the benchmark tables of Sec.~\ref{sec:lz_fit}, we
trade the original Lagrangian parameters of the Higgs sector for a set of
physically more transparent mass scales, following the conventions of
Ref.~\cite{Meng:2024lmi}.

\begin{itemize}
\item $m_A$: The characteristic mass scale of the heavy MSSM-like doublet CP-odd Higgs state, defined by
\begin{eqnarray}
m_A^2 = \left [ \lambda v_s (\sqrt{2} A_\lambda + \kappa v_s + \sqrt{2} \mu^\prime ) + 2 m_3^2 \right ]/\sin 2 \beta. \label{mA}
\end{eqnarray}    
To comply with LHC searches for heavy resonances, we maintain $m_A \sim 3\text{ TeV}$.
\item $m_B$ and $m_C$: The unmixed diagonal mass parameters of the CP-even and CP-odd singlet-dominated scalar bosons, respectively, expressed as
\begin{eqnarray}
m_B^2 &=& \frac{(A_\lambda + \mu^\prime) \sin 2 \beta}{2 \sqrt{2} v_s} \lambda v_{\rm EW}^2   + \frac{\kappa v_s}{\sqrt{2}} (A_\kappa +  2 \sqrt{2} \kappa v_s + 3 \mu^\prime ) \nonumber \\
& & - \frac{\mu}{\sqrt{2} v_s} \lambda v_{\rm EW}^2 - \frac{\sqrt{2}}{v_s} \xi^\prime.  
\end{eqnarray}
\begin{eqnarray}
m_C^2 &=&\frac{(A_\lambda + 2 \sqrt{2} \kappa v_s + \mu^\prime ) \sin 2 \beta }{2 \sqrt{2} v_s} \lambda v_{\rm EW}^2  - \frac{\kappa v_s}{\sqrt{2}} (3 A_\kappa + \mu^\prime)  \nonumber \\
& & - \frac{\mu}{\sqrt{2} v_s} \lambda v_{\rm EW}^2 - 2 m_S^{\prime\ 2} - \frac{\sqrt{2}}{v_s} \xi^\prime. \label{mC}
\end{eqnarray} 
In the small-$\lambda$ regime, these correspond closely to the physical masses of the singlet-like scalars $h_s$ and $A_s$.
\item $\widetilde{\delta}$: A dimensionless cancellation measure entering the CP-even Higgs mixing matrix element, defined via $\mathcal{M}_{S, 23}^2 \equiv \sqrt{2}\lambda \widetilde{\delta} v \mu_{\rm tot}$ with 
\begin{eqnarray}
\widetilde{\delta} \equiv \frac{2 \mu_{\rm tot} - (A_\lambda + m_N) \sin 2 \beta}{2 \mu_{\rm tot}}\label{delta}. 
\end{eqnarray}
The advantage of this parametrization is that small $\widetilde{\delta}$ allows for larger $\lambda$ while remaining consistent with the LHC measurements of the 125 GeV Higgs boson properties~\cite{Lian:2024smg}.
\item $V_{i}^{j}$: The elements of the rotation matrix to diagonalize the CP-even Higgs mass matrix, representing the $j$ field component in mass eigenstate $i$. 
\item $m_N$: The Singlino mass parameter given by $m_N = \mu^\prime + \sqrt{2} \kappa v_s$. 
\end{itemize}

\subsection{The Neutralino Sector}
\label{sec:neutralino_sector}

\subsubsection{Neutralino Mass Matrix}

The mixing among the neutral Higgsinos, Gauginos, and Singlino gives
rise to five neutralino mass eigenstates, denoted by
$\widetilde{\chi}_i^0$ ($i=1,\ldots,5$). In the gauge-eigenstate basis
\begin{equation}
    \psi^0 =
    \left(
        -i\widetilde{B},\,
        -i\widetilde{W}^0,\,
        \widetilde{H}_d^0,\,
        \widetilde{H}_u^0,\,
        \widetilde{S}
    \right),
\end{equation}
the symmetric neutralino mass matrix is given by~\cite{Ellwanger:2009dp}
\begin{equation}
\mathcal{M}_{\widetilde{\chi}^0} =
\begin{pmatrix}
 M_1 &
 0 &
 -m_Z s_W c_\beta &
 m_Z s_W s_\beta &
 0
 \\[2mm]
 0 &
 M_2 &
 m_Z c_W c_\beta &
 -m_Z c_W s_\beta &
 0
 \\[2mm]
 -m_Z s_W c_\beta &
 m_Z c_W c_\beta &
 0 &
 -\mu_{\rm tot} &
 -\dfrac{\lambda v_{\rm EW}}{\sqrt{2}}s_\beta
 \\[2mm]
 m_Z s_W s_\beta &
 -m_Z c_W s_\beta &
 -\mu_{\rm tot} &
 0 &
 -\dfrac{\lambda v_{\rm EW}}{\sqrt{2}}c_\beta
 \\[2mm]
 0 &
 0 &
 -\dfrac{\lambda v_{\rm EW}}{\sqrt{2}}s_\beta &
 -\dfrac{\lambda v_{\rm EW}}{\sqrt{2}}c_\beta &
 -(1+d)\mu_{\rm tot}
\end{pmatrix},
\label{eq:neutralino_mass_matrix}
\end{equation}
where $s_W\equiv\sin\theta_W$, $c_W\equiv\cos\theta_W$,
$s_\beta\equiv\sin\beta$, and $c_\beta\equiv\cos\beta$. 
The Singlino mass is parameterized as~\cite{Yue:2025dqe}
\begin{equation}
    m_N=-(1+d)\mu_{\rm tot},
    \qquad d>0.
\end{equation}

We adopt the convention in which the neutralino mass eigenvalues are
allowed to carry signs. Consequently, the neutralino mixing matrix
$N$ can be chosen to be real and orthogonal~\cite{Gunion:1984yn}:
\begin{equation}
    N\mathcal{M}_{\widetilde{\chi}^0}N^{T}
    =
    {\rm diag}
    \left(
        m_{\widetilde{\chi}_1^0},
        \ldots,
        m_{\widetilde{\chi}_5^0}
    \right).
\end{equation}
The corresponding mass eigenstates are
\begin{equation}
    \widetilde{\chi}_i^0
    =
    N_{i1}\psi_1^0
    +N_{i2}\psi_2^0
    +N_{i3}\psi_3^0
    +N_{i4}\psi_4^0
    +N_{i5}\psi_5^0 .
\end{equation}
The physical neutralino masses are $|m_{\widetilde{\chi}_i^0}|$, and the
Higgsino fraction of $\widetilde{\chi}_i^0$ is
$N_{i3}^2+N_{i4}^2$. We identify the lightest neutralino as a
Higgsino-dominated DM candidate when
\begin{equation}
    N_{13}^2+N_{14}^2>0.5.
\end{equation}

\subsubsection{Approximation of Neutralino Masses}

In studying the Higgsino DM, we assume $|M_1|,\,|M_2|\gg\mu_{\rm tot}\gg m_Z $ and focus on the following parameter region~\cite{Yue:2025dqe}:
\begin{equation}
    \mu_{\rm tot}\gtrsim 600~{\rm GeV},
    \quad
    \tan\beta\gtrsim20,
    \quad
    \lambda\sim0.1,
    \quad
    d\sim 0.04.
\label{eq:parameter_region_neutralino}
\end{equation}
In this regime, the three neutralinos associated predominantly with
the Higgsino--Singlino sector have signed eigenvalues close to
$-\mu_{\rm tot}$, $+\mu_{\rm tot}$, and
$-(1+d)\mu_{\rm tot}$. We therefore write
\begin{equation}
\begin{aligned}
    m_{\widetilde{\chi}_1^0}&=-\mu_{\rm tot}+\delta_1,\\
    m_{\widetilde{\chi}_2^0}&=\phantom{-}\mu_{\rm tot}+\delta_2,\\
    m_{\widetilde{\chi}_3^0}&=-\mu_{\rm tot}+\delta_3,
\end{aligned}
\qquad
|\delta_i|\ll\mu_{\rm tot}.
\label{eq:neutralino_eigenvalue_expansion}
\end{equation}
Here, the $\delta_i$ denote corrections to the signed eigenvalues and
should not be confused with the physical splitting
\begin{equation}
    \delta
    \equiv
    m_{\widetilde{\chi}_2^0}^{\rm phys}
    -
    m_{\widetilde{\chi}_1^0}^{\rm phys}
    =
    |m_{\widetilde{\chi}_2^0}|
    -
    |m_{\widetilde{\chi}_1^0}|.
\label{eq:physical_mass_splitting}
\end{equation}

Solving the characteristic equation of
Eq.~\eqref{eq:neutralino_mass_matrix} and retaining the dominant
terms gives
\begin{equation}
    \delta_2=\frac{c_1}{b_1},
    \qquad
    a\delta_{1,3}^{\,2}+b\delta_{1,3}+c=0,
\label{eq:neutralino_delta_equations}
\end{equation}
where
\begin{align}
    a={}&2\mu_{\rm tot}M_a^2,
\nonumber\\
    b={}&
    \mu_{\rm tot}m_Z^2M_c
    +2d\mu_{\rm tot}^2M_a^2
    +\frac{\lambda^2v_{\rm EW}^2}{2}
    \left(
        2M_a^2+\mu_{\rm tot}^2-M_1M_2
    \right),
\nonumber\\
    b_1={}&-4\mu_{\rm tot}^2M_b^2,
\nonumber\\
    c={}&
    d\mu_{\rm tot}^2m_Z^2M_c
    -\frac{\lambda^2\mu_{\rm tot}v_{\rm EW}^2}{2}M_a^2,
\nonumber\\
    c_1={}&
    -2\mu_{\rm tot}^2m_Z^2
    \left(M_c-2\mu_{\rm tot}\right)
    +\frac{\lambda^2\mu_{\rm tot}v_{\rm EW}^2}{2}M_b^2,
\label{eq:neutralino_coefficients}
\end{align}
with
\begin{align}
    M_a^2 &=
    M_1M_2+\mu_{\rm tot}M_1
    +\mu_{\rm tot}M_2+\mu_{\rm tot}^2,
\nonumber\\
    M_b^2 &=
    M_1M_2-\mu_{\rm tot}M_1
    -\mu_{\rm tot}M_2+\mu_{\rm tot}^2,
\nonumber\\
    M_c &=
    M_1\cos^2\theta_W
    +M_2\sin^2\theta_W
    +\mu_{\rm tot}.
\label{eq:neutralino_auxiliary_parameters}
\end{align}

Eqs.~\eqref{eq:neutralino_delta_equations}--%
\eqref{eq:neutralino_auxiliary_parameters} display separately the
two sources of the neutral-Higgsino mass splitting. The terms
proportional to $m_Z^2$ originate from Higgsino--Gaugino mixing and
reproduce the conventional MSSM contribution in the limit
$\lambda\to0$. The terms proportional to $\lambda^2v_{\rm EW}^2$, together
with the dependence on $d$, arise from Higgsino--Singlino mixing.

As $|M_1|$ and $|M_2|$ increase, the Gaugino-induced contribution
decouples, whereas the Singlino-induced contribution remains finite
and is controlled primarily by $\lambda$, $d$, and
$\mu_{\rm tot}$~\cite{Yue:2025dqe}. Therefore, even though $\lambda$ and $d$ are
numerically moderate, their contribution to the splitting can
become comparable to, or larger than, the MSSM contribution for
multi-TeV Gaugino masses. The two contributions may also interfere
constructively or destructively, allowing a physical splitting much
smaller than either characteristic contribution separately.

\subsubsection{Higgsino-Singlino Mixing}

The above observations do not imply sizable mixing between the heavy
Gauginos and the light Higgsino--Singlino states. For
$|M_{1,2}|\gg\mu_{\rm tot}$, the Gaugino components of the light
neutralinos are suppressed parametrically by
$m_Z/|M_{1,2}|$. The Gauginos therefore have only a small effect on
the composition of these states, and the full $5\times5$ matrix can
be reduced to an effective $3\times3$ Higgsino--Singlino matrix.
Nevertheless, their residual contribution can remain relevant to
the small mass splitting because the latter is a difference between
two masses of order $\mu_{\rm tot}$. Thus, a correction negligible
on the scale of the neutralino masses need not be negligible on the
scale of $\delta$, which is approximately six-order smaller than the former.  

To make the Higgsino--Singlino structure explicit, we introduce~\cite{Yue:2025dqe}
\begin{equation}
    \widetilde{H}_1
    \equiv
    \frac{\widetilde{H}_d^0+\widetilde{H}_u^0}{\sqrt{2}},
    \qquad
    \widetilde{H}_2
    \equiv
    \frac{\widetilde{H}_d^0-\widetilde{H}_u^0}{\sqrt{2}}.
\end{equation}
After neglecting the small Gaugino admixtures, the effective mass
matrix in the basis
$(\widetilde{H}_1,\widetilde{H}_2,\widetilde{S})$ becomes
\begin{equation}
\mathcal{M}_{\widetilde{\chi}^0}^{(3)}
=
\begin{pmatrix}
 -\mu_{\rm tot}
 &
 0
 &
 -\dfrac{\lambda v_{\rm EW}}{2}
 \left(s_\beta + c_\beta\right)
 \\[2mm]
 0
 &
 \mu_{\rm tot}
 &
 -\dfrac{\lambda v_{\rm EW}}{2}
 \left(s_\beta-c_\beta\right)
 \\[2mm]
 -\dfrac{\lambda v_{\rm EW}}{2}
 \left(s_\beta+c_\beta\right)
 &
 -\dfrac{\lambda v_{\rm EW}}{2}
 \left(s_\beta-c_\beta\right)
 &
 -(1+d)\mu_{\rm tot}
\end{pmatrix}.
\label{eq:reduced_neutralino_matrix}
\end{equation}
The mass eigenstates can then be written as
\begin{equation}
    \widetilde{\chi}_i^0
    =
    N'_{i1}\widetilde{H}_1
    +N'_{i2}\widetilde{H}_2
    +N'_{i3}\widetilde{S},
\label{eq:neutralino_reduced_eigenstates}
\end{equation}
where
\begin{equation}
    N'_{i1}=\frac{N_{i3}+N_{i4}}{\sqrt{2}},
    \qquad
    N'_{i2}=\frac{N_{i3}-N_{i4}}{\sqrt{2}},
    \qquad
    N'_{i3}=N_{i5}.  \label{Nprime}
\end{equation}

For $\mu_{\rm tot}>0$ and in the approximate hierarchy
\begin{equation}
    \mu_{\rm tot}\gg d\mu_{\rm tot}
    \gtrsim\lambda v_{\rm EW},
\end{equation}
the dominant mixing typically occurs between $\widetilde{H}_1$ and
$\widetilde{S}$. To leading order, the corresponding signed eigenvalues
are~\cite{Yue:2025dqe}
\begin{align}
    m_{\widetilde{\chi}_1^0}
    &\simeq
    -\mu_{\rm tot}
    +\frac{\lambda^2v_{\rm EW}^2(1+\sin2\beta)}
           {4d\mu_{\rm tot}},
\nonumber\\
    m_{\widetilde{\chi}_2^0}
    &\simeq
    \mu_{\rm tot},
\nonumber\\
    m_{\widetilde{\chi}_3^0}
    &\simeq
    -(1+d)\mu_{\rm tot}.
\label{eq:reduced_neutralino_eigenvalues}
\end{align}
The mixing matrix is approximately
\begin{equation}
    N'
    \simeq
    \begin{pmatrix}
        \cos\theta & 0 & \sin\theta\\
        0          & 1 & 0\\
        -\sin\theta& 0 & \cos\theta
    \end{pmatrix},
\label{eq:reduced_neutralino_mixing_matrix}
\end{equation}
with
\begin{equation}
    \tan\theta
    \simeq
    -\frac{\lambda v_{\rm EW}}
           {2d\mu_{\rm tot}}
    \left(s_\beta+c_\beta\right).
\label{eq:higgsino_singlino_mixing_angle}
\end{equation}
Neglecting the residual Gaugino contribution, the physical splitting
between the two Higgsino-like states is consequently
\begin{equation}
    \delta
    \simeq
    \frac{\lambda^2v_{\rm EW}^2(1+\sin2\beta)}
         {4d\mu_{\rm tot}}.
\label{eq:singlino_induced_splitting}
\end{equation}
This expression explicitly demonstrates that $\lambda$ and $d$ can
affect the small neutral-Higgsino splitting even when the
Gaugino admixtures of the relevant states are negligible. Particularly
when $d$ is small, both $|\tan \theta|$ and $\delta$ are significantly 
enhanced by the factor of $1/d$.    

If instead the dominant mixing occurs between $\widetilde{H}_2$ and
$\widetilde{S}$, which is realized when $m_N = (1 + d ) \mu_{\rm tot}$,
Eq.~(\ref{eq:reduced_neutralino_eigenvalues}) should be replaced by 
$\mu_{\rm tot} \to - \mu_{\rm tot}$ and $\sin \beta \to -\sin \beta$, 
while Eq.~(\ref{eq:higgsino_singlino_mixing_angle}) and Eq.~(\ref{eq:singlino_induced_splitting}) 
are corrected by  $\sin \beta \to -\sin \beta$. 
 
\subsubsection{Fine-Tuning Associated with the Neutralino Mass Splitting}

\label{GNMSSM-FT}

The preceding analysis identified two independent sources of the neutral-Higgsino
mass splitting. We now combine them to estimate the precision with which the
underlying parameters must be specified in order to reproduce a splitting in the
LZ-preferred range. Working in the basis
$(-i\widetilde{B},\,-i\widetilde{W}^0,\,\cos\theta\,\widetilde{H}^0_1+\sin\theta\,\widetilde{S},\,
\widetilde{H}^0_2,\,-\sin\theta\,\widetilde{H}^0_1+\cos\theta\,\widetilde{S})$  for the branch
$m_N=-(1+d)\mu_{\rm tot}$ with $\mu_{\rm tot}>0$ and $d>0$,
and treating the residual Gaugino admixtures in the mass-insertion approximation,
the leading contributions to the physical splitting take the form
\begin{equation}
\delta \simeq \left| \frac{m_Z^2}{2}
\left( \frac{s_W^2}{M_1} + \frac{c_W^2}{M_2} \right)
\left( 1 + \cos^2\theta + \sin 2\beta \sin^2\theta \right)
- \frac{\lambda^2 v_{\rm EW}^2 \left( 1 + \sin 2\beta \right)}{4 d\, \mu_{\rm tot}} \right| ,
\label{eq:delta_total}
\end{equation}
where the first term is the Gaugino-induced contribution, modulated by the
Higgsino--Singlino mixing angle $\theta$, and the second term is the
Singlino-induced contribution. Eq.~\eqref{eq:delta_total} holds for
$|M_{1,2}| \gg \mu_{\rm tot}$ and for a sizable Higgsino--Singlino mixing;
the neglected terms comprise higher-order mixing effects and radiative
corrections.

The omitted radiative corrections deserve emphasis, since they are not merely a
refinement of Eq.~\eqref{eq:delta_total}. Because the corrections to the
neutralino mass matrix are unsuppressed by the top-quark Yukawa
coupling~\cite{Chatterjee:2026scv}, and because the relevant values of $d$ are only of
order a few percent, such corrections can shift the effective value of $d$
substantially and thereby significantly alter the Singlino-induced contribution. For this
reason, we use Eq.~\eqref{eq:delta_total} only to estimate the required degree of
cancellation, and rely on a numerical treatment for all quantitative statements.

At this level of accuracy, the degree of fine tuning may be estimated by
comparing the individual contributions with the observed splitting,
\begin{equation}
\widehat{\Delta}_{\rm FT} \sim
\mathrm{Max}\left(
\left| m_Z^2 \left( \frac{s_W^2}{M_1} + \frac{c_W^2}{M_2} \right) \right| ,\;
\left| \frac{\lambda^2 v_{\rm EW}^2 \left( 1 + \sin 2\beta \right)}{4 d\, \mu_{\rm tot}} \right|
\right) \Big/ (350~{\rm keV}) .
\label{eq:FT_estimate}
\end{equation}
For multi-TeV Gauginos of equal sign, or of opposite sign but away from the
cancellation condition $M_1 c_W^2 + M_2 s_W^2 \approx 0$, the first entry in
Eq.~\eqref{eq:FT_estimate} alone yields a Gaugino-induced contribution of order 
$\mathcal{O}(1\text{ GeV})$, exceeding the LZ-preferred window by three to four orders of
magnitude. Reproducing a sub-MeV splitting therefore requires
$\widehat{\Delta}_{\rm FT} \gtrsim 10^3$ in either framework, and the small
splitting must arise from a cancellation among individually larger terms rather
than from an accidentally small contribution.

This conclusion, however, does not distinguish the two frameworks; what
distinguishes them is where the required cancellation resides. In the MSSM, a
pure-Higgsino point reproducing $\delta \simeq 350~{\rm keV}$ with $M_1 M_2 < 0$
and $|M_{1,2}|$ of a few TeV requires the two Gaugino masses to be correlated to
a relative accuracy of order $10^{-4}$. Since these same parameters fix the
Gaugino spectrum and enter the renormalization-group evolution of the soft
sector, such a correlation cannot be imposed in isolation. In the GNMSSM, by
contrast, the condition on the physical splitting no longer constrains a relation
between $M_1$ and $M_2$ alone: for a given pair of Gaugino masses, the singlet
sector supplies an additional handle with which the required residual can be
obtained. Viable configurations then exist for both signs of $M_1 M_2$,
without demanding that the Bino and Wino contributions cancel against each other.

The residual sensitivity is correspondingly carried by $\lambda$ and $d$. These
parameters are themselves subject to phenomenological constraints, as they
control the Singlino admixture $N_{15}^2$, the relic abundance, and the inelastic
cross section $\sigma_n$, motivating their detailed examination in the subsequent 
subsections. However, they do not uniquely determine the Gaugino mass spectrum 
or the unified boundary conditions at the grand unification scale. The improvement 
offered by the GNMSSM is therefore best characterized as a relocation of the required
cancellation, from a sector tightly entangled with the ultraviolet structure of
the theory to one that is comparatively isolated. We regard this reinterpretation,
rather than any reduction in the overall degree of tuning, as the principal
structural advantage of the framework.

Quantitative statements of the fine tuning require going beyond Eq.~\eqref{eq:delta_total}.
We therefore include the radiative corrections to the neutralino mass matrix and
diagonalize it numerically with high precision using
\texttt{SPheno-4.0.5}~\cite{Porod:2003um,Staub:2017jnp}. The sensitivity of the
splitting to an input parameter $p_i$ ($p_{1,2,3,4} = M_1,\, M_2,\, \lambda,\, d$) 
and the overall measure of the fine tuning are quantified by 
\begin{equation}
\Delta_{p_i} \equiv \left| \frac{\partial \ln \delta}{\partial \ln p_i} \right|, \quad \quad 
\Delta_{\rm FT}  \equiv \mathrm{Max} \left ( \Delta_{M_1}, \Delta_{M_2}, 
\Delta_{\lambda}, \Delta_{d} \right ), 
\label{eq:FT_measure}
\end{equation}
evaluated with all remaining inputs held fixed. A large $\Delta_{p_i}$ indicates 
that $\delta$ is obtained through a cancellation involving $p_i$, which must then 
be specified to a relative
accuracy of roughly $1/\Delta_{p_i}$. Beyond quantifying the overall tuning, this
measure allows the MSSM and GNMSSM realizations to be compared on a common
footing, by identifying whether the dominant sensitivity resides in parameters
tightly correlated with the remainder of the theory, such as $M_1$ and $M_2$, or
in parameters that are comparatively free, such as $\lambda$ and $d$. The values
reported for our benchmark points in Sec.~\ref{sec:lz_fit} are interpreted in
this light.

\subsection{Inelastic Neutralino-Nucleon Scattering}
\label{sec:inelastic_scattering}

The anomalous high-recoil event observed by the LZ experiment can be naturally accommodated via endothermic inelastic scattering of the lightest neutralino off target nuclei,  mediated by $t$-channel $Z$-boson exchange.
In the GNMSSM, the relevant neutral-current interaction Lagrangian in terms of the gauge eigenstates is given by
\begin{equation}
    \mathcal{L} \supset \frac{g}{4\cos\theta_W} Z_\mu \left( \overline{\widetilde{H}}_d^0 \gamma^\mu \gamma_5 \widetilde{H}_d^0 - \overline{\widetilde{H}}_u^0 \gamma^\mu \gamma_5 \widetilde{H}_u^0 \right).
    \label{eq:L_Z_gauge}
\end{equation}
Working in the convention where the neutralino mass eigenvalues can take negative values, the diagonalizing transformation $N_{ij}$ is a real orthogonal matrix. The transition to the physical Majorana mass eigenstates $\widetilde{\chi}_i^0$ yields the interaction
\begin{equation}
    \mathcal{L}_{Z\widetilde{\chi}\widetilde{\chi}} = \frac{g}{4\cos\theta_W} Z_\mu \, \overline{\widetilde{\chi}}_i^0 \gamma^\mu \left( C_{ij}^V + C_{ij}^A \gamma_5 \right) \widetilde{\chi}_j^0,
    \label{eq:L_Z_mass}
\end{equation}
where the vector and axial-vector couplings are determined by the relative signs of the mass eigenvalues $\eta_i \equiv \text{sgn}(m_{\widetilde{\chi}_i^0})$:
\begin{align}
    C_{ij}^V &= \frac{1}{2} (1 - \eta_i \eta_j) \left( N_{i3} N_{j3} - N_{i4} N_{j4} \right), \\
    C_{ij}^A &= \frac{1}{2} (1 + \eta_i \eta_j) \left( N_{i3} N_{j3} - N_{i4} N_{j4} \right).
\end{align}
As established in Sec.~\ref{sec:neutralino_sector}, for multi-TeV Gaugino masses $\left(|M_1|, |M_2| \gg |\mu_{\rm tot}|\right)$, the two lightest neutralinos originate predominantly from the Higgsino sector with opposite mass eigenvalues, $m_{\widetilde{\chi}_1^0} \simeq -\mu_{\rm tot}$ and $m_{\widetilde{\chi}_2^0} \simeq +\mu_{\rm tot}$ ($\eta_1 \eta_2 = -1$). Consequently, the diagonal $Z\widetilde{\chi}_1^0\widetilde{\chi}_1^0$ coupling is strictly axial-vector (leading to vanishing SI scattering), while the off-diagonal $Z\widetilde{\chi}_1^0\widetilde{\chi}_2^0$ interaction is purely \emph{vector-like}:
\begin{equation}
    C_{12}^V = N_{13} N_{23} - N_{14} N_{24}, \qquad C_{12}^A = 0.
\end{equation}
Under the simplified $3\times 3$ Higgsino--Singlino rotation matrix $N'$ defined in Eq.~(\ref{Nprime}) and Eq.~(\ref{eq:reduced_neutralino_mixing_matrix}), the off-diagonal coupling simplifies directly to
\begin{equation}
    C_{12}^V \simeq \cos\theta,
    \label{eq:C12_cos}
\end{equation}
where the mixing angle $\theta$ is parameterized by Eq.~(\ref{eq:reduced_neutralino_mixing_matrix}). Hence, the effective $Z\widetilde{\chi}_1^0\widetilde{\chi}_2^0$ interaction Lagrangian takes the concise form:
\begin{equation}
    \mathcal{L}_{Z\widetilde{\chi}_1^0\widetilde{\chi}_2^0} \simeq \frac{g}{4\cos\theta_W} \cos\theta \, Z_\mu \left( \overline{\widetilde{\chi}}_1^0 \gamma^\mu \widetilde{\chi}_2^0 + \overline{\widetilde{\chi}}_2^0 \gamma^\mu \widetilde{\chi}_1^0 \right).
    \label{eq:L_Z_eff_GNMSSM}
\end{equation}

At low momentum transfer ($q^2 \ll m_Z^2$), integrating out the $Z$ boson generates the effective four-fermion interaction with the Standard Model quarks $q \in \{u, d, s, c, b, t\}$:
\begin{equation}
    \mathcal{L}_{\rm eff} = -\frac{G_F}{\sqrt{2}} \, \cos\theta \, \left( \overline{\widetilde{\chi}}_2^0 \gamma^\mu \widetilde{\chi}_1^0 \right) \sum_q \overline{q} \gamma_\mu \left( v_q - a_q \gamma_5 \right) q,
\end{equation}
where $G_F/\sqrt{2} = g^2 / (8 m_Z^2 \cos^2\theta_W)$.
In the non-relativistic regime, the quark vector currents match coherently onto the nucleon currents $\langle N | \overline{q} \gamma_\mu q | N \rangle = f_q^N \overline{u}_N \gamma_\mu u_N$. Summing over the valence quark content of protons and neutrons yields the effective neutralino--nucleon vector couplings:
\begin{equation}
    \mathcal{L}_{\rm eff}^N = -\frac{G_F}{\sqrt{2}} \, \left( \overline{\widetilde{\chi}}_2^0 \gamma^\mu \widetilde{\chi}_1^0 \right) \left[ f_p \, \overline{p} \gamma_\mu p + f_n \, \overline{n} \gamma_\mu n \right],
\end{equation}
with the explicit forms of $f_p$ and $f_n$ given by
\begin{align}
    f_p &= \cos\theta \left( 2 v_u + v_d \right) = \cos\theta \left( \frac{1}{2} - 2\sin^2\theta_W \right), \label{eq:fp_def} \\
    f_n &= \cos\theta \left( v_u + 2 v_d \right) = -\frac{1}{2}\cos\theta. \label{eq:fn_def}
\end{align}

Therefore, the DM--proton coupling is substantially suppressed compared to the DM--neutron coupling, $|f_p / f_n| \simeq 0.08$. The coherent nuclear scattering is thus overwhelmingly dominated by the neutron content of the nucleus.

Matching these microscopic effective couplings onto the differential inelastic cross section in Eq.~(\ref{eq:dsigmadER}),
we can uniquely identify the zero-momentum-transfer reference inelastic scattering cross section per neutron $\sigma_n$. In the limit of pure Higgsino DM ($\cos\theta \to 1$), as realized in the MSSM with heavy decoupled Gauginos, $\sigma_n$ reduces to
\begin{equation}
    \sigma_n^{\rm MSSM} = \frac{G_F^2 \mu_n^2}{2\pi} \simeq 7.4 \times 10^{-39} \text{ cm}^2,
    \label{eq:sigma_n_MSSM}
\end{equation}

In the GNMSSM, the non-negligible Singlino component in $\widetilde{\chi}_1^0$ (with Singlino fraction $N_{15}^2 \simeq \sin^2\theta \lesssim 18\%$) dilutes the active Higgsino component. Consequently, the inelastic cross section is structurally suppressed by a factor of $\cos^2\theta$:
\begin{equation}
    \sigma_n^{\rm GNMSSM} \simeq \cos^2\theta \, \sigma_n^{\rm MSSM} = \left( 1 - \sin^2\theta \right) \frac{G_F^2 \mu_n^2}{2\pi}.
    \label{eq:sigma_n_GNMSSM}
\end{equation}

For our benchmark points, presented in Sec.~\ref{sec:lz_fit}, this suppression yields reference cross sections in the range $\sigma_n^{\rm GNMSSM} \simeq (6.0 - 7.2) \times 10^{-39} \text{ cm}^2$. This demonstrates an important advantage of the GNMSSM over the MSSM: the inelastic scattering 
rate is no longer rigidly tied to the pure gauge coupling, providing the possibility for the Higgsino explanation to reconcile the single high-recoil event with terrestrial exclusion limits and astrophysical capture rates.

\subsection{Thermal relic abundance}
\label{subsec:relic_density}

For completeness, we briefly summarize the mechanisms controlling the
thermal relic abundance in the parameter region considered here.
Higgsino--Singlino mixing modifies the couplings of
\(\widetilde{\chi}_1^0\) and therefore changes its annihilation and
coannihilation cross sections relative to those of a pure Higgsino.
This allows the observed relic abundance to be obtained for a
DM mass in the window $m_{\widetilde{\chi}_1^0} \sim 600\text{--}1100$~GeV, 
significantly below the canonical pure-Higgsino value of
approximately \(1.1~{\rm TeV}\)~\cite{Yue:2025dqe}.

Nearly degenerate Sleptons may provide an additional thermal effect~\cite{Chakraborti:2017dpu}.
We parameterize their common soft-mass scale by
\begin{equation}
    (m_{\widetilde L}^2)_{ij}
    =
    (m_{\widetilde e}^2)_{ij}
    =
    \left[(1+\Delta_L)\mu_{\rm tot}\right]^2\delta_{ij}.
    \label{eq:DeltaL_definition}
\end{equation}
Thus, \(\Delta_L\) characterizes the relative separation between the
Slepton and Higgsino mass scales. The thermally averaged effective
annihilation cross section is~\cite{Griest:1990kh,Baker:2015qna}
\begin{equation}
    \langle\sigma_{\rm eff}v\rangle
    =
    \sum_{ij}
    \langle\sigma_{ij}v\rangle
    \frac{g_ig_j}{g_{\rm eff}^2}
    (1+\Delta_i)^{3/2}
    (1+\Delta_j)^{3/2}
    e^{-x(\Delta_i+\Delta_j)},
    \label{eq:sigma_eff}
\end{equation}
where $\sigma_{ij}$ represents the cross section of the annihilation $\widetilde{\chi}_i \widetilde{\chi}_j \to X Y$, $g_i$ denotes the degrees of freedom for species $i$, and 
\begin{equation}
    \Delta_i
    =
    \frac{m_{\widetilde{\chi}_i}-m_{\widetilde{\chi}_1^0}}
         {m_{\widetilde{\chi}_1^0}},
    \qquad
    g_{\rm eff}
    =
    \sum_i
    g_i(1+\Delta_i)^{3/2}e^{-x\Delta_i}.
    \label{eq:geff}
\end{equation}
In the benchmark scenarios, the Sleptons do not necessarily provide
the dominant annihilation channels. Their presence nevertheless
changes \(g_{\rm eff}\) and hence the effective freeze-out rate~\cite{Chakraborti:2017dpu}. The
observed relic abundance therefore results from the combined effects
of Higgsino--Singlino mixing and, for sufficiently small
\(\Delta_L\), the thermal population of nearby Slepton states. The
relic density used below is calculated numerically with all relevant
annihilation and coannihilation channels included.
The combined effect of these two mechanisms allows the Higgsino mass 
scale to be as low as $\sim 300\,\mathrm{GeV}$ while still producing 
the observed relic abundance. 

\section{\label{numerical results}Numerical Results}
\label{sec:lz_fit}

The analytical discussion of Sec.~\ref{sec:gnmssm} identifies the ingredients
that a GNMSSM explanation of the LZ event must combine: a Higgsino-dominated
lightest neutralino, a Singlino admixture that adjusts both the mass splitting
$\delta$ and the inelastic cross section $\sigma_n^\text{GNMSSM}$, multi-TeV Gaugino masses
of either relative sign, and a relic abundance obtained through modified
(co)annihilation. In this section we examine whether these ingredients can be
realized simultaneously in explicit parameter points that also satisfy all
other experimental constraints. To this end we draw on the comprehensive
numerical study of Higgsino DM in the GNMSSM that we performed in
Ref.~\cite{Yue:2025dqe}, and we briefly recall the computational setup adopted there.
Particle spectra were computed with \texttt{SPheno}-4.0.5~\cite{Porod:2003um,Staub:2017jnp},
interfaced with \texttt{FlavorKit} for flavor observables; DM observables were
evaluated with \texttt{micrOMEGAs}-5.0.4~\cite{Belanger:2001fz,Belanger:2006is,Belanger:2008sj,Belanger:2013oya,Alguero:2023zol};
the properties of the $125~\mathrm{GeV}$ Higgs boson were confronted with data
through global fits performed with
\texttt{HiggsSignals}-2.6.2~\cite{HS2013xfa,HSConstraining2013hwa,HS2014ewa,HS2020uwn} and
\texttt{HiggsTools}-1.2~\cite{Bahl:2022igd}, while the additional Higgs states were
constrained by \texttt{HiggsBounds}-5.10.2~\cite{HB2008jh,HB2011sb,HBHS2012lvg,HB2013wla,HB2020pkv} 
using the LEP, Tevatron, and LHC exclusion limits; indirect-detection
constraints were imposed with the \texttt{MADHAT} 
package~\cite{Boddy:2018qur, Boddy:2019kuw, Boddy:2024tiu}, based on 14-year Fermi-LAT observations
of 54 dwarf spheroidal galaxies; and the flavor observables
$Br(B_s\to\mu^+\mu^-)$ and $Br(B\to X_s\gamma)$ were required to agree with their
measured values at the $2\sigma$ level. We emphasize that the neutralino mass
matrix includes the radiative corrections discussed in Sec.~\ref{GNMSSM-FT}
and is diagonalized numerically. This treatment is indispensable here, because
the $\mathcal{O}(100)~\mathrm{keV}$ splitting relevant to the LZ event is far
below the size of the loop corrections to the individual neutralino masses.

\begin{table}[thb]
\centering
\resizebox{1\textwidth}{!}
{
\begin{tabular}{crcr|crcr}
\hline \hline
\multicolumn{4}{c|}{\bf Benchmark Point BP1 } & \multicolumn{4}{c}{\bf Benchmark Point BP2} \\ \hline
$\lambda$ & 0.053& $m_h$ & 125.3~GeV& $\lambda$ & 0.050 & $m_h$ & 125.2~GeV\\
$\kappa$ & -0.239& $m_{A_s}$ & 387.6~GeV& $\kappa$ & 0.117 & $m_{A_s}$ & 390.2~GeV\\
$\widetilde\delta$& -0.161& $m_{\widetilde{\chi}_1^0}$ & -662.170457~GeV& $\widetilde\delta$& -0.150 & $m_{\widetilde{\chi}_1^0}$ & 772.685316~GeV\\
$d$ & 0.054& $m_{\widetilde{\chi}_2^0}$ & 662.170790~GeV& $d$ & 0.068 & $m_{\widetilde{\chi}_2^0}$ & -772.685657~GeV\\
$\Delta_L$ & 0.141& $\bm\delta$&\textbf{333.0~keV}& $\Delta_L$ & 0.116 &$\bm\delta$& \textbf{341.0~keV}\\
$\tan\beta$ & 29.797& $m_{\widetilde{\chi}_3^0}$ & -684.1~GeV& $\tan\beta$ & 27.674 & $m_{\widetilde{\chi}_3^0}$ & -809.4~GeV\\
$\mu$ & 625.7~GeV& $m_{\widetilde{\chi}_4^0}$ & 2738.7~GeV& $\mu$ & 735.9~GeV & $m_{\widetilde{\chi}_4^0}$ & 2975.5~GeV\\
$v_s$ & 600.0~GeV & $m_{\widetilde{\chi}_5^0}$ & 3168.9~GeV& $v_s$ & 600.0~GeV & $m_{\widetilde{\chi}_5^0}$ & -3483.3~GeV\\
$\mu_{\rm tot}$ & 648.3~GeV& $m_{\widetilde{\chi}_1^\pm}$ & 663.5~GeV& $\mu_{\rm tot}$ & 757.1~GeV & $m_{\widetilde{\chi}_1^\pm}$ & 773.3~GeV\\
$A_\lambda$ & 23143.3~GeV& $m_{\widetilde{\chi}_2^\pm}$ & 3169.0~GeV& $A_\lambda$ & 24929.4~GeV & $m_{\widetilde{\chi}_2^\pm}$ & 2975.5~GeV\\
$A_\kappa$ & 1562.2~GeV& $m_{\widetilde l_1}$ & 665.8~GeV& $A_\kappa$ & 3344.0~GeV & $m_{\widetilde l_1}$ & 781.1~GeV\\
$A_t/A_b$ & 2770.0~GeV& $m_{\widetilde l_2}$ & 671.8~GeV& $A_t/A_b$ & 2684.1~GeV & $m_{\widetilde l_2}$ & 795.9~GeV\\
$M_1$ & 2741.4~GeV& $m_{\widetilde l_3}$ & 671.9~GeV& $M_1$ & -3477.7~GeV & $m_{\widetilde l_3}$ & 796.1~GeV\\
$M_2$ & 3200.3~GeV& $m_{\widetilde l_4}$ & 720.0~GeV& $M_2$ & 3002.7~GeV & $m_{\widetilde l_4}$ & 807.3~GeV\\
$m_A$ & 3000.0~GeV & $m_{\widetilde l_5}$ & 720.1~GeV& $m_A$ & 3000.0~GeV & $m_{\widetilde l_5}$ & 807.5~GeV\\
$m_B$ & 171.4~GeV& $m_{\widetilde l_6}$ & 738.3~GeV& $m_B$ & 202.5~GeV & $m_{\widetilde l_6}$ & 831.6~GeV\\
$m_C$ & 400.0~GeV & $m_{\widetilde \nu_1}$ & 667.2~GeV& $m_C$ & 400.0~GeV & $m_{\widetilde \nu_1}$ & 792.1~GeV\\
$m_{N}$& -683.3~GeV& $m_{\widetilde \nu_2}$ & 667.2~GeV& $m_{N}$& -808.7~GeV& $m_{\widetilde \nu_2}$ & 792.1~GeV\\
$m_{h_s}$& 189.7~GeV& $m_{\widetilde \nu_3}$ & 671.9~GeV& $m_{h_s}$& 189.8~GeV& $m_{\widetilde \nu_3}$ & 795.5~GeV\\ \hline
\multicolumn{2}{c}{$V_{h_s}^S,~V_{h_s}^{\rm SM},~V_{h}^S,~V_{h}^{\rm SM}$} & \multicolumn{2}{c|}{~-0.983,~~0.185,~-0.185,~-0.983} & \multicolumn{2}{c}{$V_{h_s}^S,~V_{h_s}^{\rm SM},~V_{h}^S,~V_{h}^{\rm SM}$} & \multicolumn{2}{c}{~-0.993,~~0.116,~-0.116,~-0.993}\\
\multicolumn{2}{c}{$N_{11},~N_{12},~N_{13},~N_{14},~N_{15}$} & \multicolumn{2}{c|}{~0.008, ~-0.013, ~-0.669, ~-0.671, ~~0.319
} & \multicolumn{2}{c}{$N_{11},~N_{12},~N_{13},~N_{14},~N_{15}$} & \multicolumn{2}{c}{~0.008, ~~0.025, ~-0.707, ~~0.706, ~~0.004}\\
\multicolumn{2}{c}{$N_{21},~N_{22},~N_{23},~N_{24},~N_{25}$} & \multicolumn{2}{c|}{~0.015, ~-0.022,~~ 0.708,~-0.706, ~-0.005
} & \multicolumn{2}{c}{$N_{21},~N_{22},~N_{23},~N_{24},~N_{25}$} & \multicolumn{2}{c}{-0.011, ~-0.014, ~-0.696, ~-0.697, ~~0.172}\\
\multicolumn{2}{c}{$N_{31},~N_{32},~N_{33},~N_{34},~N_{35}$} & \multicolumn{2}{c|}{-0.003, ~~0.004, ~~0.229,~~0.223, ~~0.948
} & \multicolumn{2}{c}{$N_{31},~N_{32},~N_{33},~N_{34},~N_{35}$} & \multicolumn{2}{c}{~0.002, ~~0.002, ~~0.124, ~~0.119, ~~0.985}\\
\multicolumn{2}{c}{$N_{41},~N_{42},~N_{43},~N_{44},~N_{45}$} & \multicolumn{2}{c|}{~0.999, ~~0.004, ~-0.005, ~~0.017, ~~0.000
} & \multicolumn{2}{c}{$N_{41},~N_{42},~N_{43},~N_{44},~N_{45}$} & \multicolumn{2}{c}{~0.000, ~-0.996, ~-0.008, ~~0.028, ~~0.000}\\
\multicolumn{2}{c}{$N_{51},~N_{52},~N_{53},~N_{54},~N_{55}$} & \multicolumn{2}{c|}{~0.003, ~-0.999, ~-0.006, ~~0.026, ~~0.000
} & \multicolumn{2}{c}{$N_{51},~N_{52},~N_{53},~N_{54},~N_{55}$} & \multicolumn{2}{c}{~0.999, ~-0.000, ~-0.003, ~-0.013, ~~0.000}\\ \hline
\multicolumn{2}{c}{$\Omega h^2$}                                                                                  & \multicolumn{2}{c|}{0.131} & \multicolumn{2}{c}{$\Omega h^2$}                                                                                  & \multicolumn{2}{c}{0.118}                                                                                  \\ 
\multicolumn{2}{c}{$\sigma^{\rm SI}_{\rm eff}$}                                                                                  & \multicolumn{2}{c|}{$1.08\times 10^{-47}~{\rm cm}^2$} & \multicolumn{2}{c}{$\sigma^{\rm SI}_{\rm eff}$}                                                                                  & \multicolumn{2}{c}{$1.05\times 10^{-47}~{\rm cm}^2$}                                                                                  \\ 
\multicolumn{2}{c}{$\sigma^{\rm SD}_n$}                                                                                  & \multicolumn{2}{c|}{$4.86\times 10^{-43}~{\rm cm}^2$} & \multicolumn{2}{c}{$\sigma^{\rm SD}_n$}                                                                                  & \multicolumn{2}{c}{$4.17\times 10^{-44}~{\rm cm}^2$}                                                                                  \\ 
\multicolumn{2}{c}{$\sigma_n^\text{GNMSSM}$}                                                                                  & \multicolumn{2}{c|}{$6.65\times 10^{-39}~{\rm cm}^2$} & \multicolumn{2}{c}{$\sigma_n^\text{GNMSSM}$}                                                                                  & \multicolumn{2}{c}{$7.2\times 10^{-39}~{\rm cm}^2$}                                                                                  \\ 
\multicolumn{2}{c}{$\Delta\chi^2$}                                                                                  & \multicolumn{2}{c|}{0.258} & \multicolumn{2}{c}{$\Delta\chi^2$}                                                                                  & \multicolumn{2}{c}{0.36}                                                                                  \\ 
\hline
\multicolumn{2}{c}{ Annihilations }                                                                                  & \multicolumn{2}{c|}{Fractions [\%]} & \multicolumn{2}{c}{Annihilations}                                                                                  & \multicolumn{2}{c}{Fractions [\%]}                                                                                  \\
\multicolumn{2}{c}{$\widetilde{\chi}_2^0\widetilde{\chi}_1^- \to  d\bar u/s\bar c \cdots$} & \multicolumn{2}{l|}{6.5~/~6.5 $\cdots$}        & \multicolumn{2}{c}{$\widetilde{\chi}_1^0\widetilde{\chi}_1^- \to  d\bar u/s\bar c \cdots$} & \multicolumn{2}{l}{7.0~/~7.0 $\cdots$}        \\ 
\multicolumn{2}{c}{$\widetilde{\chi}_1^0\widetilde{\chi}_1^- \to  d\bar u/s\bar c \cdots$} & \multicolumn{2}{l|}{5.8~/~5.8 $\cdots$}        & \multicolumn{2}{c}{$\widetilde{\chi}_2^0\widetilde{\chi}_1^- \to  d\bar u/s\bar c \cdots$} & \multicolumn{2}{l}{6.8~/~6.8 $\cdots$}        \\ 
\multicolumn{2}{c}{$\widetilde{\chi}_1^0\widetilde{\chi}_2^0 \to  d\bar d/s\bar s \cdots$} & \multicolumn{2}{l|}{3.1~/~3.1 $\cdots$}        & \multicolumn{2}{c}{$\widetilde{\chi}_1^0\widetilde{\chi}_2^0 \to  t\bar t \cdots$} & \multicolumn{2}{l}{3.7 $\cdots$}        \\ 
\multicolumn{2}{c}{$\cdots$} & \multicolumn{2}{l|}{$\cdots$}        & \multicolumn{2}{c}{$\cdots$} & \multicolumn{2}{l}{$\cdots$}      \\
\hline
\multicolumn{2}{c}{$\Delta_{\rm FT}$}                                                                                   & \multicolumn{2}{c|}{$\Delta_{M_1,M_2,\lambda,d} \simeq 2728, 6970, 38000, 364$} & \multicolumn{2}{c}{$\Delta_{\rm FT}$}                                                                                  & \multicolumn{2}{c}{$\Delta_{M_1,M_2,\lambda,d} \simeq 1906, 7481, 2200, 40$}                                                                                  \\ 
 \hline \hline
\end{tabular}}
\caption{\label{tab:BP12} Benchmark points BP1 and BP2: detailed parameter specifications consistent with all current experimental constraints, characterized by a lower Higgsino mass regime ($600$--$800\,\rm{GeV}$) and TeV-scale gaugino masses $M_1$ and $M_2$. In particular, the two lightest neutralino masses $m_{\widetilde{\chi}^0_{1,2}}$ are presented to six-digit precision to resolve the $\mathcal{O}(100)\,\mathrm{keV}$ mass splitting $\delta$ required to accommodate the LZ 248\,keV nuclear recoil event. The effective spin-independent cross section is defined as
$\sigma^{\rm SI}_{\rm eff} = 0.169\,\sigma^{\rm SI}_{p} + 0.347\,\sigma^{\rm SI}_{n}
+ 0.484\sqrt{\sigma^{\rm SI}_{p}\sigma^{\rm SI}_{n}}$~\cite{Cao:2019aam}, where
$\sigma^{\rm SI}_{p}$ and $\sigma^{\rm SI}_{n}$ denote the spin-independent cross
sections on protons and neutrons, respectively, and the coefficients account for the
natural abundances of the xenon isotopes~\cite{XENON:2018voc}. Also listed are the
neutron spin-dependent cross section $\sigma^{\rm SD}_{n}$ and the inelastic
scattering cross section $\sigma^{\rm GNMSSM}_{n}$. The fine-tuning measure
$\Delta_{\rm FT}$ is defined in Eq.~(\ref{eq:FT_measure}). Furthermore, an extended likelihood analysis yields $\Delta\chi^2 = 0.258$ (BP1) and $\Delta\chi^2 = 0.36$ (BP2) relative to the global best-fit point, demonstrating excellent agreement with the LZ excess. Finally, numbers following each annihilation channel indicate its fractional contribution to the total DM annihilation cross section at the freeze-out temperature. }
\end{table}

\begin{table}[th]
\centering
\resizebox{1\textwidth}{!}
{
\begin{tabular}{crcr|crcr}
\hline \hline
\multicolumn{4}{c|}{\bf Benchmark Point BP3} & \multicolumn{4}{c}{\bf Benchmark Point BP4} \\ \hline
$\lambda$ & 0.050& $m_h$ & 125.1~GeV& $\lambda$ & 0.070& $m_h$ & 124.8~GeV\\
$\kappa$ & 0.021& $m_{A_s}$ & 404.2~GeV& $\kappa$ & 0.092& $m_{A_s}$ & 382.0~GeV\\
$\widetilde\delta$&  -0.181& $m_{\widetilde{\chi}_1^0}$ & -837.451151~GeV& $\widetilde\delta$& -0.119& $m_{\widetilde{\chi}_1^0}$ & -901.352954~GeV\\
$d$ & 0.051& $m_{\widetilde{\chi}_2^0}$ & 837.451493~GeV& $d$ & 0.068& $m_{\widetilde{\chi}_2^0}$ & 901.353295~GeV\\
$\Delta_L$ & 0.212& $\bm\delta$&\textbf{342.0~keV}& $\Delta_L$ & 0.152&$\bm \delta $& \textbf{341.0~keV}\\
$\tan\beta$ & 27.937& $m_{\widetilde{\chi}_3^0}$ & -864.0~GeV& $\tan\beta$ & 19.550& $m_{\widetilde{\chi}_3^0}$ & -946.7~GeV\\
$\mu$ & 799.6~GeV& $m_{\widetilde{\chi}_4^0}$ & 3915.7~GeV& $\mu$ & 855.2~GeV& $m_{\widetilde{\chi}_4^0}$ & 3346.9~GeV\\
$v_s$ & 600.0~GeV& $m_{\widetilde{\chi}_5^0}$ & 4324.5~GeV& $v_s$ & 600.0~GeV & $m_{\widetilde{\chi}_5^0}$ & 3688.4~GeV\\
$\mu_{\rm tot}$ & 820.7~GeV& $m_{\widetilde{\chi}_1^\pm}$ & 838.5~GeV& $\mu_{\rm tot}$ & 884.9~GeV& $m_{\widetilde{\chi}_1^\pm}$ & 902.5~GeV\\
$A_\lambda$ & 27983.3~GeV& $m_{\widetilde{\chi}_2^\pm}$ & 3915.7~GeV& $A_\lambda$ & 20349.1~GeV& $m_{\widetilde{\chi}_2^\pm}$ & 3688.5~GeV\\
$A_\kappa$ & 6184.4~GeV& $m_{\widetilde l_1}$ & 872.0~GeV& $A_\kappa$ & 4182.4~GeV& $m_{\widetilde l_1}$ & 932.0~GeV\\
$A_t/A_b$ & 2906.0~GeV& $m_{\widetilde l_2}$ & 877.2~GeV& $A_t/A_b$ & 2965.4~GeV& $m_{\widetilde l_2}$ & 934.9~GeV\\
$M_1$ & 4303.3~GeV& $m_{\widetilde l_3}$ & 877.2~GeV& $M_1$ & 3344.3~GeV& $m_{\widetilde l_3}$ & 934.9~GeV\\
$M_2$ & 3963.2~GeV& $m_{\widetilde l_4}$ & 933.2~GeV& $M_2$ & 3731.6~GeV& $m_{\widetilde l_4}$ & 993.2~GeV\\
$m_A$ & 3000.0~GeV & $m_{\widetilde l_5}$ & 933.2~GeV& $m_A$ & 3000.0~GeV & $m_{\widetilde l_5}$ & 993.2~GeV\\
$m_B$ & 181.2~GeV& $m_{\widetilde l_6}$ & 946.3~GeV& $m_B$ & 224.0~GeV& $m_{\widetilde l_6}$ & 999.6~GeV\\
$m_C$ & 400.0~GeV & $m_{\widetilde \nu_1}$ & 873.6~GeV& $m_C$ & 400.0~GeV & $m_{\widetilde \nu_1}$ & 931.5~GeV\\
$m_{N}$& -862.6~GeV& $m_{\widetilde \nu_2}$ & 873.6~GeV& $m_{N}$& -961.0~GeV& $m_{\widetilde \nu_2}$ & 931.5~GeV\\
$m_{h_s}$& 183.8~GeV& $m_{\widetilde \nu_3}$ & 876.7~GeV& $m_{h_s}$& 189.8~GeV& $m_{\widetilde \nu_3}$ & 932.9~GeV\\ \hline
\multicolumn{2}{c}{$V_{h_s}^S,~V_{h_s}^{\rm SM},~V_{h}^S,~V_{h}^{\rm SM}$} & \multicolumn{2}{c|}{-0.987,~~0.160,~--0.160,~-0.987} & \multicolumn{2}{c}{$V_{h_s}^S,~V_{h_s}^{\rm SM},~V_{h}^S,~V_{h}^{\rm SM}$} & \multicolumn{2}{c}{-0.987,~~0.160,~--0.160,~-0.987}\\
\multicolumn{2}{c}{$N_{11},~N_{12},~N_{13},~N_{14},~N_{15}$} & \multicolumn{2}{c|}{~0.006, ~-0.011, ~-0.685, ~-0.687, ~~0.241} & \multicolumn{2}{c}{$N_{11},~N_{12},~N_{13},~N_{14},~N_{15}$} & \multicolumn{2}{c}{~0.007, ~-0.011, ~-0.692, ~-0.694, ~~0.200}\\
\multicolumn{2}{c}{$N_{21},~N_{22},~N_{23},~N_{24},~N_{25}$} & \multicolumn{2}{c|}{~0.009, ~-0.018, ~~0.707, ~-0.707, ~-0.003} & \multicolumn{2}{c}{$N_{21},~N_{22},~N_{23},~N_{24},~N_{25}$} & \multicolumn{2}{c}{~0.013, ~-0.020, ~~0.707, ~-0.706, ~-0.004}\\
\multicolumn{2}{c}{$N_{31},~N_{32},~N_{33},~N_{34},~N_{35}$} & \multicolumn{2}{c|}{-0.001, ~~0.003, ~~0.173, ~~0.168, ~~0.971} & \multicolumn{2}{c}{$N_{31},~N_{32},~N_{33},~N_{34},~N_{35}$} & \multicolumn{2}{c}{-0.001, ~~0.002, ~~0.144, ~~0.138, ~~0.980}\\
\multicolumn{2}{c}{$N_{41},~N_{42},~N_{43},~N_{44},~N_{45}$} & \multicolumn{2}{c|}{~0.002, ~~0.999, ~~0.005, ~-0.021, ~~0.000} & \multicolumn{2}{c}{$N_{41},~N_{42},~N_{43},~N_{44},~N_{45}$} & \multicolumn{2}{c}{~~0.999, ~~0.004, ~-0.004, ~~0.014, ~~0.000}\\
\multicolumn{2}{c}{$N_{51},~N_{52},~N_{53},~N_{54},~N_{55}$} & \multicolumn{2}{c|}{-0.999, ~~0.002, ~~0.002, ~-0.011, ~~0.000} & \multicolumn{2}{c}{$N_{51},~N_{52},~N_{53},~N_{54},~N_{55}$} & \multicolumn{2}{c}{~0.004, ~-0.999, ~-0.006, ~~0.022, ~~0.000}\\ \hline
\multicolumn{2}{c}{$\Omega h^2$}                                                                                  & \multicolumn{2}{c|}{0.100} & \multicolumn{2}{c}{$\Omega h^2$}                                                                                  & \multicolumn{2}{c}{0.119}                                                                                  \\ 
\multicolumn{2}{c}{$\sigma^{\rm SI}_{\rm eff}$}                                                                                  & \multicolumn{2}{c|}{$1.98\times 10^{-49}~{\rm cm}^2$} & \multicolumn{2}{c}{$\sigma^{\rm SI}_{\rm eff}$}                                                                                  & \multicolumn{2}{c}{$1.13\times 10^{-48}~{\rm cm}^2$}                                                                                  \\ 
\multicolumn{2}{c}{$\sigma^{\rm SD}_n$}                                                                                  & \multicolumn{2}{c|}{$1.55\times 10^{-43}~{\rm cm}^2$} & \multicolumn{2}{c}{$\sigma^{\rm SD}_n$}                                                                                  & \multicolumn{2}{c}{$1.76\times 10^{-43}~{\rm cm}^2$}                                                                                  \\ 
\multicolumn{2}{c}{$\sigma_n^\text{GNMSSM}$}                                                                                  & \multicolumn{2}{c|}{$6.97\times 10^{-39}~{\rm cm}^2$} & \multicolumn{2}{c}{$\sigma_n^\text{GNMSSM}$}                                                                                  & \multicolumn{2}{c}{$7.10\times 10^{-39}~{\rm cm}^2$}                                                                                  \\ 
\multicolumn{2}{c}{$\Delta\chi^2$}                                                                                  & \multicolumn{2}{c|}{0.05} & \multicolumn{2}{c}{$\Delta\chi^2$}                                                                                  & \multicolumn{2}{c}{0.473}                                                                                  \\ 
\hline
\multicolumn{2}{c}{ Annihilations }                                                                                  & \multicolumn{2}{c|}{Fractions [\%]} & \multicolumn{2}{c}{Annihilations}                                                                                  & \multicolumn{2}{c}{Fractions [\%]}                                                                                  \\
\multicolumn{2}{c}{$\widetilde{\chi}_2^0\widetilde{\chi}_1^- \to  d\bar u/s\bar c/ b\bar{t} \cdots $} & \multicolumn{2}{l|}{6.9~/~6.9~/~3.4 $\cdots$}        & \multicolumn{2}{c}{$\widetilde{\chi}_1^0\widetilde{\chi}_2^0 \to t\bar t \cdots$} & \multicolumn{2}{l}{7.7 $\cdots$}        \\ 
\multicolumn{2}{c}{$\widetilde{\chi}_1^0\widetilde{\chi}_1^- \to  d\bar u/s\bar c/b\bar{t} \cdots$} & \multicolumn{2}{l|}{6.5~/~6.5~/~3.6 $\cdots$}        & \multicolumn{2}{c}{$\widetilde{\chi}_2^0\widetilde{\chi}_1^- \to  d\bar u/s\bar c \cdots $} & \multicolumn{2}{l}{6.5~/~6.5 $\cdots$}        \\ 
\multicolumn{2}{c}{$\widetilde{\chi}_1^0\widetilde{\chi}_2^0 \to  t\bar t \cdots$} & \multicolumn{2}{l|}{5.9 $\cdots$}        & \multicolumn{2}{c}{$\widetilde{\chi}_1^0\widetilde{\chi}_1^- \to d\bar{u}/s\bar{c} \cdots$} & \multicolumn{2}{l}{6.3~/~6.3 $\cdots$}        \\ 
\multicolumn{2}{c}{$\cdots$} & \multicolumn{2}{l|}{$\cdots$}        & \multicolumn{2}{c}{$\cdots$} & \multicolumn{2}{l}{$\cdots$}        \\ 
\hline
\multicolumn{2}{c}{$\Delta_{\rm FT}$}                                                                                   & \multicolumn{2}{c|}{$\Delta_{M_1,M_2,\lambda,d} \simeq 1613, 5571, 5718, 6$} & \multicolumn{2}{c}{$\Delta_{\rm FT}$}                                                                                  & \multicolumn{2}{c}{$\Delta_{M_1,M_2,\lambda,d} \simeq 2199, 6157, 1319, 1173$}                                                                                     \\ 
 \hline \hline
\end{tabular}}
\caption{Benchmark points BP3 and BP4 feature an intermediate Higgsino mass regime ($800$--$900\,\rm{GeV}$); other specifications as in Table~\ref{tab:BP12}.}
\label{tab:BP34}
\end{table}

\begin{table}[th]
\centering
\resizebox{1\textwidth}{!}
{
\begin{tabular}{crcr|crcr}
\hline \hline
\multicolumn{4}{c|}{\bf Benchmark Point BP5} & \multicolumn{4}{c}{\bf Benchmark Point BP6} \\ \hline
$\lambda$ & 0.054 & $m_h$ & 125.3~GeV& $\lambda$ & 0.056& $m_h$ & 124.9~GeV\\
$\kappa$ & 0.124 & $m_{A_s}$ & 388.6~GeV& $\kappa$ & -0.105& $m_{A_s}$ & 419.3~GeV\\
$\widetilde\delta$&  -0.050& $m_{\widetilde{\chi}_1^0}$ & 984.818389~GeV& $\widetilde\delta$& -0.150& $m_{\widetilde{\chi}_1^0}$ & -1107.90794~GeV\\
$d$ & 0.056& $m_{\widetilde{\chi}_2^0}$ & -984.818738~GeV& $d$ & 0.028& $m_{\widetilde{\chi}_2^0}$ & 1107.90829~GeV\\
$\Delta_L$ & 0.179& $\bm\delta$&\textbf{349.0~keV}& $\Delta_L$ & 0.201&$\bm \delta $& \textbf{350.0~keV}\\
$\tan\beta$ & 23.504& $m_{\widetilde{\chi}_3^0}$ & -1021.7~GeV& $\tan\beta$ & 26.426& $m_{\widetilde{\chi}_3^0}$ & -1126.0~GeV\\
$\mu$ & 944.1~GeV& $m_{\widetilde{\chi}_4^0}$ & 4022.8~GeV& $\mu$ & 1068.7~GeV& $m_{\widetilde{\chi}_4^0}$ & 2281.4~GeV\\
$v_s$ & 600.0~GeV& $m_{\widetilde{\chi}_5^0}$ & 4661.0~GeV& $v_s$ & 600.0~GeV & $m_{\widetilde{\chi}_5^0}$ & 4979.4~GeV\\
$\mu_{\rm tot}$ & 966.8~GeV& $m_{\widetilde{\chi}_1^\pm}$ & 985.7~GeV& $\mu_{\rm tot}$ & 1092.4~GeV& $m_{\widetilde{\chi}_1^\pm}$ & 1109.1~GeV\\
$A_\lambda$ & 24920.4~GeV& $m_{\widetilde{\chi}_2^\pm}$ & 4022.9~GeV& $A_\lambda$ & 34376.7~GeV& $m_{\widetilde{\chi}_2^\pm}$ & 2281.6~GeV\\
$A_\kappa$ & 3714.5~GeV& $m_{\widetilde l_1}$ & 1024.0~GeV& $A_\kappa$ & 2765.3~GeV& $m_{\widetilde l_1}$ & 1240.7~GeV\\
$A_t/A_b$ & 3876.9~GeV& $m_{\widetilde l_2}$ & 1029.4~GeV& $A_t/A_b$ & 3012.1~GeV& $m_{\widetilde l_2}$ & 1244.8~GeV\\
$M_1$ & 4632.5~GeV& $m_{\widetilde l_3}$ & 1029.5~GeV& $M_1$ & 4943.5~GeV& $m_{\widetilde l_3}$ & 1244.8~GeV\\
$M_2$ & 4072.8~GeV& $m_{\widetilde l_4}$ & 1074.7~GeV& $M_2$ & 2283.4~GeV& $m_{\widetilde l_4}$ & 1296.9~GeV\\
$m_A$ & 3000.0~GeV & $m_{\widetilde l_5}$ & 1074.7~GeV& $m_A$ & 3000.0~GeV & $m_{\widetilde l_5}$ & 1296.9~GeV\\
$m_B$ & 170.2~GeV& $m_{\widetilde l_6}$ & 1085.0~GeV& $m_B$ & 153.7~GeV& $m_{\widetilde l_6}$ & 1306.0~GeV\\
$m_C$ & 400.0~GeV & $m_{\widetilde \nu_1}$ & 1026.4~GeV& $m_C$ & 400.0~GeV & $m_{\widetilde \nu_1}$ & 1294.3~GeV\\
$m_{N}$& -1020.9~GeV& $m_{\widetilde \nu_2}$ & 1026.4~GeV& $m_{N}$& -1123.0~GeV& $m_{\widetilde \nu_2}$ & 1294.3~GeV\\
$m_{h_s}$& 153.9~GeV& $m_{\widetilde \nu_3}$ & 1028.3~GeV& $m_{h_s}$& 190.4~GeV& $m_{\widetilde \nu_3}$ & 1296.0~GeV\\ \hline
\multicolumn{2}{c}{$V_{h_s}^S,~V_{h_s}^{\rm SM},~V_{h}^S,~V_{h}^{\rm SM}$} & \multicolumn{2}{c|}{-0.977,~~0.214,~--0.214,~-0.977} & \multicolumn{2}{c}{$V_{h_s}^S,~V_{h_s}^{\rm SM},~V_{h}^S,~V_{h}^{\rm SM}$} & \multicolumn{2}{c}{-0.989,~~0.146,~-0.146,~-0.989}\\
\multicolumn{2}{c}{$N_{11},~N_{12},~N_{13},~N_{14},~N_{15}$} & \multicolumn{2}{c|}{~0.009, ~-0.018, ~~0.707, ~-0.707, ~-0.003} & \multicolumn{2}{c}{$N_{11},~N_{12},~N_{13},~N_{14},~N_{15}$} & \multicolumn{2}{c}{-0.004, ~~0.014, ~~0.638, ~~0.640, ~-0.428}\\
\multicolumn{2}{c}{$N_{21},~N_{22},~N_{23},~N_{24},~N_{25}$} & \multicolumn{2}{c|}{~0.005, ~-0.010, ~-0.694, ~-0.695, ~~0.185} & \multicolumn{2}{c}{$N_{21},~N_{22},~N_{23},~N_{24},~N_{25}$} & \multicolumn{2}{c}{-0.008, ~~0.047, ~-0.707, ~~0.705, ~~0.003}\\
\multicolumn{2}{c}{$N_{31},~N_{32},~N_{33},~N_{34},~N_{35}$} & \multicolumn{2}{c|}{-0.001, ~~0.002, ~~0.133, ~~0.129, ~~0.983} & \multicolumn{2}{c}{$N_{31},~N_{32},~N_{33},~N_{34},~N_{35}$} & \multicolumn{2}{c}{~0.002, ~-0.007, ~-0.305, ~-0.301, ~~0.904}\\
\multicolumn{2}{c}{$N_{41},~N_{42},~N_{43},~N_{44},~N_{45}$} & \multicolumn{2}{c|}{~0.002, ~~0.999, ~~0.006, ~-0.020, ~~0.000} & \multicolumn{2}{c}{$N_{41},~N_{42},~N_{43},~N_{44},~N_{45}$} & \multicolumn{2}{c}{~0.001, ~~0.999, ~-0.008, ~~0.028, ~~0.000}\\
\multicolumn{2}{c}{$N_{51},~N_{52},~N_{53},~N_{54},~N_{55}$} & \multicolumn{2}{c|}{-0.999, ~~0.001, ~~0.002, ~-0.010, ~~0.000} & \multicolumn{2}{c}{$N_{51},~N_{52},~N_{53},~N_{54},~N_{55}$} & \multicolumn{2}{c}{-0.999, ~~0.000, ~~0.002, ~-0.009, ~~0.000}\\ \hline
\multicolumn{2}{c}{$\Omega h^2$}                                                                                  & \multicolumn{2}{c|}{0.120} & \multicolumn{2}{c}{$\Omega h^2$}                                                                                  & \multicolumn{2}{c}{0.124}                                                                                  \\ 
\multicolumn{2}{c}{$\sigma^{\rm SI}_{\rm eff}$}                                                                                  & \multicolumn{2}{c|}{$6.56\times 10^{-48}~{\rm cm}^2$} & \multicolumn{2}{c}{$\sigma^{\rm SI}_{\rm eff}$}                                                                                  & \multicolumn{2}{c}{$6.53\times 10^{-48}~{\rm cm}^2$}                                                                                  \\ 
\multicolumn{2}{c}{$\sigma^{\rm SD}_n$}                                                                                  & \multicolumn{2}{c|}{$2.42\times 10^{-44}~{\rm cm}^2$} & \multicolumn{2}{c}{$\sigma^{\rm SD}_n$}                                                                                  & \multicolumn{2}{c}{$2.48\times 10^{-43}~{\rm cm}^2$}                                                                                  \\ 
\multicolumn{2}{c}{$\sigma_n^\text{GNMSSM}$}                                                                                  & \multicolumn{2}{c|}{$7.15\times 10^{-39}~{\rm cm}^2$} & \multicolumn{2}{c}{$\sigma_n^\text{GNMSSM}$}                                                                                  & \multicolumn{2}{c}{$6.04\times 10^{-39}~{\rm cm}^2$}                                                                                  \\ 
\multicolumn{2}{c}{$\Delta\chi^2$}                                                                                  & \multicolumn{2}{c|}{0.164} & \multicolumn{2}{c}{$\Delta\chi^2$}                                                                                  & \multicolumn{2}{c}{0.03}                                                                                  \\ 
\hline
\multicolumn{2}{c}{ Annihilations }                                                                                  & \multicolumn{2}{c|}{Fractions [\%]} & \multicolumn{2}{c}{Annihilations}                                                                                  & \multicolumn{2}{c}{Fractions [\%]}                                                                                  \\
\multicolumn{2}{c}{$\widetilde{\chi}_1^0\widetilde{\chi}_2^0 \to  t\bar t \cdots$} & \multicolumn{2}{l|}{10.5 $\cdots$}        & \multicolumn{2}{c}{$\widetilde{\chi}_1^0\widetilde{\chi}_2^0 \to t\bar t \cdots$} & \multicolumn{2}{l}{10.5 $\cdots$}        \\ 
\multicolumn{2}{c}{$\widetilde{\chi}_1^0\widetilde{\chi}_1^- \to  d\bar u/s\bar c/b\bar{t} \cdots$} & \multicolumn{2}{l|}{6.0~/~6.0~/~3.8 $\cdots$}        & \multicolumn{2}{c}{$\widetilde{\chi}_1^+\widetilde{\chi}_1^- \to  t\bar t/b\bar{b} \cdots$} & \multicolumn{2}{l}{6.2~/~3.2 $\cdots$}        \\ 
\multicolumn{2}{c}{$\widetilde{\chi}_2^0\widetilde{\chi}_1^- \to  d\bar u/s\bar c/ b\bar{t} \cdots $} & \multicolumn{2}{l|}{5.8~/~5.8~/~4.3 $\cdots$}        & \multicolumn{2}{c}{$\widetilde{\chi}_2^0\widetilde{\chi}_1^- \to d\bar{u}/s\bar{c}/b\bar{t} \cdots$} & \multicolumn{2}{l}{5.8~/~5.8~/~5.4 $\cdots$}        \\ 
\multicolumn{2}{c}{$\widetilde{\chi}_1^+\widetilde{\chi}_1^- \to  t\bar t/ b\bar b \cdots$} & \multicolumn{2}{l|}{5.6 ~/~2.3 $\cdots$}        & \multicolumn{2}{c}{$\widetilde{\chi}_1^0\widetilde{\chi}_1^- \to b\bar t/d\bar{u}/s\bar c \cdots$} & \multicolumn{2}{l}{5.0~/~4.7~/~4.7 $\cdots$}        \\ 
\multicolumn{2}{c}{$\cdots$} & \multicolumn{2}{l|}{$\cdots$}        & \multicolumn{2}{c}{$\cdots$} & \multicolumn{2}{l}{$\cdots$}        \\ 
\hline
\multicolumn{2}{c}{$\Delta_{\rm FT}$}                                                                                   & \multicolumn{2}{c|}{$\Delta_{M_1,M_2,\lambda,d} \simeq 1432, 5442, 3438, 162$} & \multicolumn{2}{c}{$\Delta_{\rm FT}$}                                                                                  & \multicolumn{2}{c}{$\Delta_{M_1,M_2,\lambda,d} \simeq 1428, 17143, 62857, 317$}                                                                                 \\ 
 \hline \hline
\end{tabular}}
\caption{Benchmark points BP5 and BP6 feature a heavy Higgsino mass regime ($900$--$1100\,\rm{GeV}$); other specifications as in Table~\ref{tab:BP12}.}
\label{tab:BP56}
\end{table}

\begin{figure}
	\centering
	\includegraphics[width=0.7\linewidth]{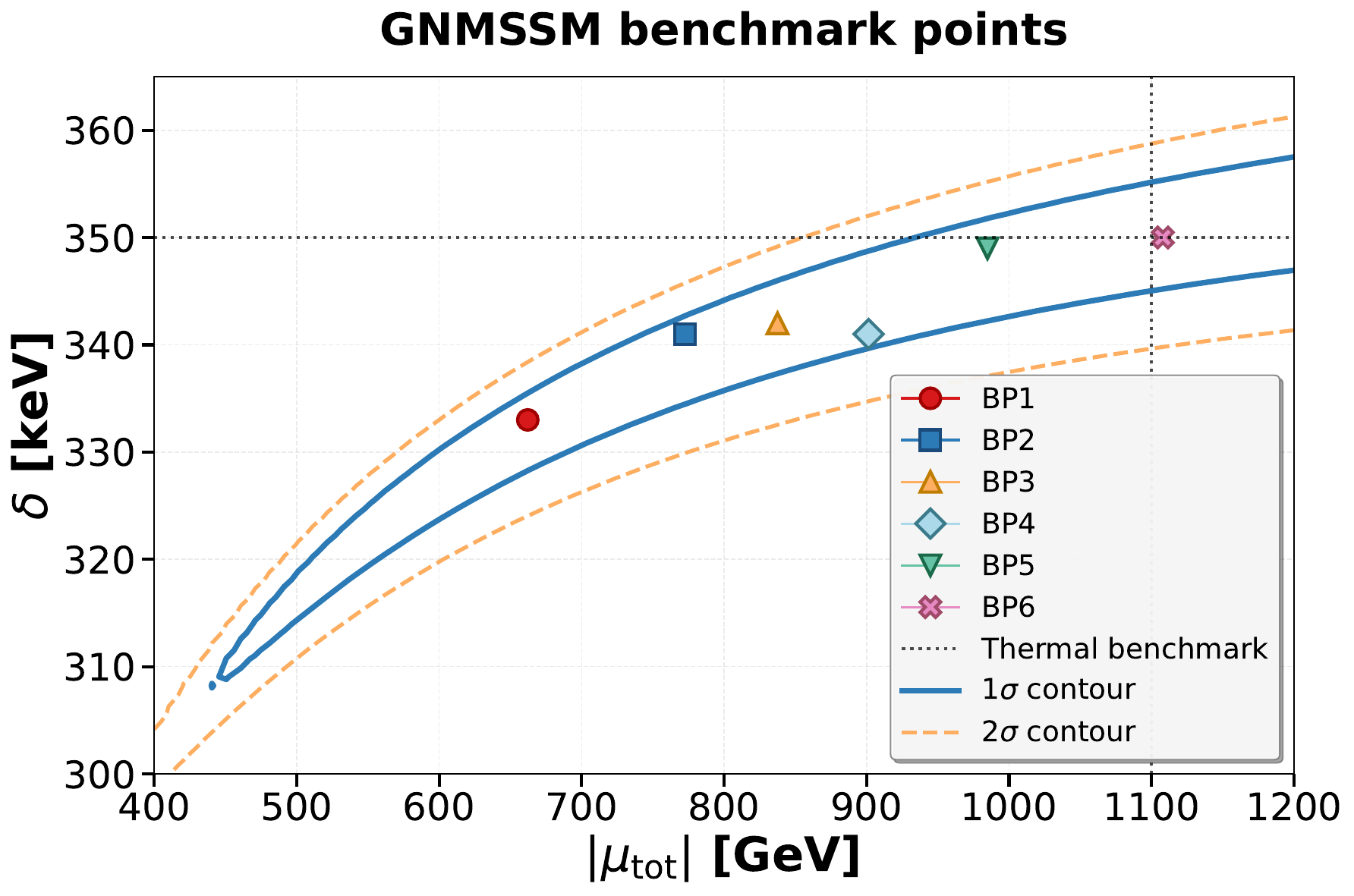}
	\caption{Exclusion contours at $1\sigma$ (solid) and $2\sigma$ (dashed) in the $(|\mu_{\mathrm{tot}}|, \, \delta)$ plane for inelastic Higgsino DM under the SHM, calibrated to the thermal benchmark $(1100 \, \mathrm{GeV}, 350 \, \mathrm{keV})$ of Ref.~\cite{Fan:2026kxx}. Six GNMSSM benchmark points (BP1--BP6) are shown with distinct markers; all lie within the $1\sigma$ contour $\left(\Delta \chi^2 <1\right)$.}
	\label{fig:bps}
\end{figure}

From the viable parameter space of Ref.~\cite{Yue:2025dqe} we select six
representative benchmark points, BP1--BP6, whose detailed properties are
listed in Tables~\ref{tab:BP12}, \ref{tab:BP34}, and~\ref{tab:BP56}. All six
points satisfy the experimental constraints enumerated above; in addition, they
are compatible with the recent LHC searches for compressed Higgsino spectra
summarized in Ref.~\cite{Su:2025wyu}. To confront them with the LZ observation, we
evaluate the extended likelihood of Sec.~\ref{sec:likelihood} in the
$(|\mu_{\rm tot}|,\delta)$ plane, assuming the Standard Halo Model and the
pure-Higgsino reference cross section
$\sigma_n^\text{MSSM} = 7.4\times 10^{-39}~\mathrm{cm}^2$, and derive the corresponding
$1\sigma$ and $2\sigma$ contours. The results are shown in Fig.~\ref{fig:bps}, 
in which the thermal Higgsino benchmark of Ref.~\cite{Fan:2026kxx}, 
$(m_{\widetilde\chi^0_1},\delta)=(1.1~\mathrm{TeV},350~\mathrm{keV})$, is indicated 
for reference. All six benchmark points lie inside the $1\sigma$
contour, i.e. $\Delta\chi^2<1$, with the best-fit point BP6 reaching
$\Delta\chi^2 = 0.03$. Since the $\Delta\chi^2$ of each point is computed with
its actual GNMSSM cross section $\sigma_n^\text{GNMSSM}\propto\cos^2\theta$ rather than with
the reference value, the agreement with the LZ event is not an artefact of the
normalization adopted for the contours.

The two lightest neutralino masses in the tables are quoted to six significant
digits so that the sub-MeV splitting $\delta = |m_{\widetilde\chi^0_2}|-|m_{\widetilde\chi^0_1}|$
can be read off directly. Together with the Gaugino masses, mixing-matrix
elements, cross sections, dominant (co)annihilation channels, and fine-tuning
measures listed alongside, this information allows the mechanisms of
Sec.~\ref{sec:gnmssm} to be traced point by point. Doing so, we find that the
benchmark points address the three rigidities of the MSSM Higgsino scenario
identified in Sec.~\ref{sec:introduction} in the following ways.

\begin{enumerate}
\item Relaxed restrictions on the Gaugino sector.
All six points realize $\delta\simeq 330$--$350~\mathrm{keV}$ with Gaugino
mass parameters of magnitude $2$--$5~\mathrm{TeV}$, and they include both
$M_1M_2>0$ (e.g. BP1, BP3--BP6) and $M_1M_2<0$ (e.g. BP2)
configurations. In each case the Singlino-induced contribution to the
splitting compensates the Gaugino-induced shift, so that neither an
intermediate Gaugino scale of order $10^7~\mathrm{GeV}$ nor the precise
Bino--Wino cancellation characteristic of the TeV-scale MSSM alternative is
needed. The small-splitting condition is thereby transferred from a
restrictive relation within the Gaugino sector to a matching condition that
involves the singlet sector as well. We stress, in line with
Sec.~\ref{GNMSSM-FT}, that this is a relocation rather than a removal of
the tuning: the fine-tuning measures in the tables exceed
$5\times 10^{3}$ for every point, with the largest sensitivity carried by
$M_2$ or $\lambda$ depending on the point. The benchmark points therefore
demonstrate increased model-building flexibility in the Gaugino sector, but
they do not establish the viability of arbitrary Gaugino mass patterns, nor the
absence of fine tuning.

\item Adjustable inelastic scattering cross section.
The Singlino admixture of the DM state, $|N_{15}|^2$, ranges from a nearly
negligible value (BP2) to the ten-percent level, and the associated mixing
factor $\cos^2\theta$ reduces the inelastic cross section from the
pure-Higgsino value to $\sigma_n^\text{GNMSSM}\simeq 6$--$7.2\times 10^{-39}~\mathrm{cm}^2$.
The suppression is modest for the points considered, but it illustrates that
in the GNMSSM the normalization of the inelastic rate is no longer fixed by
the gauge coupling alone. This freedom is absent in the MSSM, where the same
Higgsino--Gaugino mixing that sets $\delta$ also fixes the scattering rates,
so that the two cannot be varied independently.

\item Flexible DM mass. In the MSSM, the observed relic abundance is typically
accounted for by standard thermal freeze-out of a $1.1~\mathrm{TeV}$ Higgsino. 
The GNMSSM relaxes this
correlation in two ways, both of which are realized among the benchmark
points: Higgsino--Singlino mixing modifies the annihilation and
coannihilation rates into SM final states, and coannihilation with Sleptons
nearly degenerate with the DM state (cf. the Slepton masses $m_{\widetilde l_i}$
in the tables) contributes appreciably to the effective annihilation cross
section at freeze-out. As a result, the DM mass spans
$m_{\widetilde\chi^0_1}\simeq 660$--$1110~\mathrm{GeV}$ across BP1--BP6, extending
well below the MSSM value while reproducing $\Omega h^2$ within about $3\%$ of
the measured value. A lighter Higgsino is also phenomenologically welcome, as
it may alleviate the tension with the higher-energy LZ sideband that
constrains the MSSM interpretation~\cite{Rodd:2026tyn}.
\end{enumerate}

In summary, all six benchmark points are consistent with the aforementioned 
experimental constraints in explaining the LZ high-energy recoil
event. Compared with the MSSM interpretation, the GNMSSM thus provides a
framework in which the LZ event can be accommodated without an ultraheavy or
finely correlated Gaugino sector. One qualification, however, should be kept
in mind:  as quantified above and in Sec.~\ref{GNMSSM-FT}, the required sub-MeV
splitting is still obtained through a cancellation among contributions that
are individually several orders of magnitude larger; since $\delta$ is smaller
than the DM mass by more than six orders of magnitude and no symmetry is known
to enforce such a hierarchy, a comparable degree of tuning appears to be a
generic feature of inelastic-DM explanations of the LZ excess rather than a
peculiarity of the present model.

\section{Solar Neutrino Constraints and Potential Evasions}

\label{Solar-neutrino}

\subsection{Assumptions and Uncertainties in Solar Neutrino Constraints}
\label{sec:solar-astro}

The benchmark points presented in Sec.~\ref{sec:lz_fit} satisfy the
direct-detection, collider, and relic-density requirements, but they have not
yet been confronted with the constraints derived from the high-energy neutrino
searches of IceCube and Super-Kamiokande
~\cite{Pospelov:2026ewn,Bose:2026ndd,DiMauro:2026dqp,Nguyen:2026lui}. As
discussed in Sec.~\ref{sec:introduction}, these searches impose the requirement
$\delta \gtrsim 566~\mathrm{keV}$ on a pure thermal Higgsino, which is in
pronounced tension with the splitting preferred by the LZ event. Before
examining what the GNMSSM can offer in this respect, it is therefore necessary
to clarify how robust this bound actually is. As emphasized at the end of
Ref.~\cite{Bose:2026ndd}, the bound rests on several astrophysical inputs,
including the solar capture rate, the subsequent evolution of the captured population, 
and the neutrino yield from its annihilation, whose uncertainties are potentially 
sizable. We summarize them below, since they set
the stage for the model-specific mechanisms discussed in the following
subsections.

\begin{itemize}
\item Orbital evolution and planetary perturbations. For inelastic Higgsino DM, 
tree-level up-scattering inside the Sun, 
$\chi_1 N\to\chi_2 N$, is kinematically allowed only in the inner solar region where
$\tfrac{1}{2}\mu_A v^2>\delta$. For the mass splitting preferred by the LZ event, 
this condition is typically satisfied within $r \lesssim 0.2\,R_\odot$, where the 
local escape velocity exceeds $\sim 1300~{\rm km/s}$. Once the DM particle 
is captured and loses energy, the inelastic channel eventually closes, and 
any further sinking toward the solar core must proceed through much weaker 
elastic scatterings. In this situation, the efficiency of 
thermalization and the competition between orbital damping and 
gravitational perturbations from the giant planets become important. 
If thermalization is slow, a fraction of the captured DM population 
may remain on wide eccentric orbits for a long time and thus be more 
vulnerable to planetary perturbations. Existing analyses treat the
orbital damping in an idealized manner, and this is widely regarded 
as the principal loophole in the solar neutrino constraint on 
inelastic Higgsino DM.

\item Halo substructure and local velocity distribution. The LMC
(\(M\sim1.5\times10^{11}M_\odot\)) induces a reflex motion of the inner Galaxy of
\(30\)--\(40~\mathrm{km/s}\) relative to the outer halo, while cold streams such as S1 may
contribute a few percent of the local density at relative speeds up to
\(\sim 500~\mathrm{km/s}\). Because the inelastic rate is controlled by the integral above
\(v_{\min}=\sqrt{2\delta/\mu_A}\), which for \(\delta\sim350~\mathrm{keV}\) lies near
\(v_{\rm esc}\) on the exponentially falling tail, such shifts can modify the terrestrial
rate by factors of a few. Crucially, however, this effect does not significantly weaken
the solar bound: inside the Sun one has
\(v^2(r)=v_\infty^2+v_{\rm esc}^2(r)\) with \(v_{\rm esc}\gg v_\infty\), so a
\(40~\mathrm{km/s}\) shift changes \(v^2\) by less than \(5\%\), while the corrections to
\(\rho_\odot\) are at most at the ten-percent level. Non-standard halo kinematics
therefore shift the LZ-preferred window upward, while leaving the neutrino bound
essentially unchanged.

\item Subleading solar and neutrino-transport systematics. The capture 
of heavy inelastic DM in the Sun receives important contributions 
from relatively heavy elements. Consequently, uncertainties in the solar 
heavy-element abundances propagate into the predicted capture rate. 
In addition, neutrino absorption, regeneration, and energy degradation 
inside the Sun affect the conversion between the DM annihilation rate and 
the observable neutrino event rate. These effects do not invalidate 
the IceCube/Super-K limits, but they do imply that the precise numerical 
bound is somewhat model dependent.
\end{itemize}

These considerations indicate that, on the solar side, the most important
source of uncertainty is the capture-to-annihilation history, namely the
interplay between orbital evolution and thermalization inside the Sun. By
contrast, halo substructure mainly acts as an external astrophysical input that
shifts the LZ-preferred range of $\delta$ rather than directly weakening the
IceCube/Super-K bound itself. When both effects are taken into account, the
tension between the LZ-preferred parameter region and the neutrino-telescope
limits may be appreciably relaxed, as pointed out clearly 
in Ref.~\cite{Bose:2026ndd}. Nevertheless, for the minimal thermal 
Higgsino scenario, these constraints still represent a potentially
serious challenge. This motivates the question of whether the GNMSSM itself,
beyond the model-independent astrophysical considerations above, provides
additional means of suppressing the solar neutrino signal.

\subsection{Suppression of Neutrino Signals through a Singlet Resonance}
\label{sec:solar-funnel}

The uncertainties discussed in Sec.~\ref{sec:solar-astro} are common to any
inelastic Higgsino scenario and do not depend on the underlying supersymmetric
model. In the GNMSSM, however, the situation becomes considerably more flexible
owing to its richer structure. In addition to the suppression of the inelastic scattering rate by 
Higgsino--Singlino mixing and the decorrelation between the DM mass 
and the relic abundance, the singlet sector of the GNMSSM introduces
extra states and couplings, characterized by parameters 
such as $\kappa$, $A_\kappa$, $m_{A_s}$, and $m_{h_s}$. These 
parameters are largely independent of those governing the LZ signal 
and are only weakly constrained by current data~\cite{Meng:2024lmi}.
As we will show in this subsection, these ingredients 
allow the neutrino signal from DM annihilation in the Sun to be modified 
in a way that is not available in the MSSM, namely through the 
annihilation final states rather than through the
capture rate.

To isolate the effect of the annihilation final states, we first assume
that capture--annihilation equilibrium has been reached and that
evaporation and other losses of captured particles are negligible.
The total annihilation rate then satisfies
$\Gamma_{\rm ann}=C_{\rm cap}/2$, where $C_{\rm cap}$ denotes the effective capture rate. For a specified capture rate and
annihilation spatial distribution, the observable neutrino signal is
therefore controlled by the annihilation branching fractions and the
neutrino spectra after propagation through the Sun. Increasing the
total annihilation cross section does not by itself reduce the
equilibrium annihilation rate; instead, the mechanism considered here
requires a channel that dominates annihilation under solar conditions
while producing few neutrinos in the energy range relevant to the
searches. We accordingly seek a channel satisfying the following
two requirements:
\begin{enumerate}
    \item[\bf{(i)}] it supports $s$-wave annihilation and can remain dominant at
    the low relative velocities relevant inside the Sun;
    \item[\bf{(ii)}] its decay cascade yields relatively few high-energy neutrinos
    after propagation through the solar medium.
\end{enumerate}

A possible realization of these requirements involves a singlet-dominated 
resonance in the CP-violating GNMSSM. Consider the case in which the singlet-dominated CP-odd Higgs boson
$A_s$ is lighter than the doublet-like CP-odd state $A$ and decays predominantly
into $h_s h_s$ through the CP-violating trilinear coupling
$A_s h_s h_s\propto{\rm Im}(\kappa A_\kappa)$.\footnote{For concreteness, 
we assume that CP violation originates solely from a complex soft trilinear parameter
$A_\kappa$ with ${\rm Re}(A_\kappa)\gg{\rm Im}(A_\kappa)$. The analysis is then
simplified by treating the CP-violating term in the soft-breaking Lagrangian of
Eq.~(\ref{Soft-terms}) as an interaction, while regarding the predominantly 
CP-odd and CP-even states in the small-CP-mixing limit by the fields 
$A_s$ and $h_s$ in the  CP-conserving limit, respectively.} The DM then annihilates through the
$s$-channel process $\widetilde\chi_1^0\widetilde\chi_1^0\to A_s^\ast\to h_s h_s$, whose
amplitude is proportional to the product of the $A_s h_s h_s$ coupling and the
$\widetilde\chi_1^0\widetilde\chi_1^0 A_s$ coupling; the latter receives a contribution
$\lambda N_{13}N_{14}\simeq\lambda/2$ from the superpotential term
$\lambda\widehat S\widehat H_u\cdot\widehat H_d$ and a contribution
$\kappa N_{15}^2\simeq\kappa\sin^2\theta$ from the singlet self-interaction. If
the mediator lies close to resonance, $m_{A_s}\simeq2m_{\widetilde\chi_1^0}$, the
measured relic density can be reproduced for $m_{\widetilde\chi_1^0}\gg1~{\rm TeV}$
even for small values of these couplings, so that the DM mass is decoupled from
the value of about $1.1~{\rm TeV}$ dictated by cosmology in the pure-Higgsino
case.

We stress that the CP-violating nature of the $A_s h_s h_s$ vertex is essential
for requirement~{\bf{(i)}}. For a Majorana pair, the $s$-wave initial state is
${}^1S_0$ with $J^{PC}=0^{-+}$, whereas two identical CP-even scalars in an
$L=0$ configuration form a $0^{++}$ state. In the CP-conserving limit the
transition $\widetilde\chi_1^0\widetilde\chi_1^0\to h_s h_s$ is therefore forbidden at
$v\to0$ and proceeds only in the $p$-wave ($L=1$), which is fatal in the solar core
where $v^2\sim10^{-10}$. The CP-violating trilinear vertex circumvents 
this selection rule, allowing unsuppressed $s$-wave annihilation at $v \to 0$.

Requirement~{\bf{(ii)}} is met by the same channel once $h_s$ is sufficiently
light. Because the $h_s h_s$ mode saturates the total annihilation rate, the
branching ratios of the $WW$ and $ZZ$ final states in the solar core are
correspondingly reduced. If, in addition, $m_{h_s}\lesssim3~{\rm GeV}$, i.e.\
below both the $\tau^+\tau^-$ and the open-charm thresholds, $h_s$ decays
predominantly into $e^+e^-$, $\mu^+\mu^-$, $\pi^+\pi^-$ and light hadrons.
Provided that the singlet--doublet mixing is sufficient for $h_s$ to decay 
inside the solar interior, these light charged particles and charged pions 
are rapidly decelerated and absorbed by the dense solar core plasma 
($\rho_\odot \sim 150~\mathrm{g/cm^3}$). The resulting neutrinos, 
originating primarily from muons decaying at rest, have maximum energies 
$E_\nu \le m_\mu/2 \approx 53$~MeV (with only a negligible higher-energy
leakage from kaon decays), which lie far below the
energy range probed by the IceCube and Super-K DM searches. Consequently, 
the final neutrino yield above detector threshold can be strongly suppressed.

In this neutrino-poor regime, observable high-energy neutrino events 
arise almost entirely from the residual $W^+W^-$ and $ZZ$ modes. 
The expected event rate scales as
\begin{equation}
N_\nu \propto \frac{1}{2} C_{\rm cap} \, \mathrm{Br}_\odot(WW/ZZ) \sim \left( \frac{1.1~\mathrm{TeV}}{m_{\widetilde\chi_1^0}} \right)^{\!3} \times N_\nu^{\rm MSSM}(m_{\widetilde\chi_1^0} = 1.1~\mathrm{TeV}) ,
\label{eq:Nnu_scaling}
\end{equation}
where one power of $m_{\widetilde\chi_1^0}^{-1}$ originates from the 
geometric reduction in the solar capture rate at a fixed local DM 
mass density ($\rho_0/m_{\widetilde\chi_1^0}$), and the remaining factor 
of $m_{\widetilde\chi_1^0}^{-2}$ arises from the branching ratio into 
gauge bosons, $\mathrm{Br}_\odot(WW/ZZ) \propto \langle\sigma v\rangle_{WW} \propto m_{\widetilde\chi_1^0}^{-2}$, 
evaluated against a fixed total cross section required by freeze-out. 
We emphasize that the precise size of this suppression depends on the position of the $A_s$ pole
relative to $2m_{\widetilde\chi_1^0}$, since solar annihilation samples $v\to0$
whereas freeze-out samples $v\sim0.3$~\cite{Griest:1990kh}, and also on the modified
neutrino spectrum and the enhanced solar absorption at higher DM masses.
Eq.~(\ref{eq:Nnu_scaling}) should therefore be regarded as a parametric
estimate rather than a precise prediction.

\subsection{Pathways to Relax the Solar Neutrino Constraints in the GNMSSM}

The preceding two subsections have approached the solar neutrino constraints
from complementary directions: Sec.~\ref{sec:solar-astro} identified the
astrophysical assumptions on which the bound relies, while
Sec.~\ref{sec:solar-funnel} presented a mechanism, specific to the GNMSSM, that
acts on the annihilation final states. We now collect the factors currently
known to modify the solar neutrino constraints within the GNMSSM and, in
particular, clarify how the resonant funnel mechanism relates to the benchmark
points of Sec.~\ref{sec:lz_fit}, which do not rely on it.

Broadly speaking, the GNMSSM provides three distinct, yet complementary,
avenues to potentially relax the stringent solar neutrino constraints that confront the
minimal MSSM interpretation:
\begin{itemize}
\item Modified solar capture through mixing and mass-dependent kinematics.
The benchmark points BP1--BP6 lie in the comparatively low-mass region, 
$m_{\widetilde\chi_1^0}\simeq 600$--$1100~\mathrm{GeV}$, where Higgsino--Singlino 
mixing and optionally Slepton coannihilation allow the observed relic 
abundance to be reproduced. 
In this regime the bound is relaxed
in two ways. First, the mixing suppresses the inelastic cross section by the factor
$\cos^2\theta$ discussed in Sec.~\ref{sec:gnmssm}. Second, the lower DM mass reduces the
DM--nucleus reduced mass $\mu_A$, so that the kinematic condition
$\tfrac12\mu_A v^2>\delta$ for up-scattering is met only in a smaller fraction of the
solar interior. Both effects tend to lower the effective capture rate $C_{\rm cap}$ and hence
the equilibrium neutrino flux, without any modification of the annihilation final states.
Whether this suppression alone is sufficient can only be established by a 
dedicated calculation of the solar neutrino signal for these benchmark points.

\item Modified post-capture evolution through inefficient thermalization.
A second possibility arises because some SUSY explanations of the LZ event are
accompanied by SI and SD elastic scattering cross sections far below those adopted
in Ref.~\cite{Pospelov:2026ewn}. Such nearly vanishing rates result from cancellations between
tree-level and loop-level contributions~\cite{Wu:2026nhi}. Since the elastic channel governs
the final stage of energy loss once the inelastic channel has closed
(cf.\ the first item at the beginning of this section), its suppression renders the
thermalization of the captured population inefficient and leaves it spatially more
diffuse than the thermal profile usually assumed. This reduces the effective
annihilation rate and may drive the system away from capture--annihilation
equilibrium. The resulting relaxation of the IceCube/Super-K bound on $\delta$ has
not yet been precisely quantified, and we defer a dedicated study to future work.

\item Reduced high-energy neutrino yields through altered annihilation branching fractions.
The resonant scenario instead moves to $m_{\widetilde\chi_1^0}\gg 1.1$~TeV,
where the observable neutrino flux is suppressed according to 
Eq.~(\ref{eq:Nnu_scaling}). This route acts on the annihilation final states 
rather than on capture, and is therefore
independent of the two mechanisms above; its price is the larger electroweak
fine-tuning associated with a multi-TeV $\mu_{\rm tot}$.
It is not realized by the benchmark points of Sec.~\ref{sec:lz_fit}, 
but represents an alternative region of the GNMSSM parameter space 
that deserves exploration in its own right.
\end{itemize}

Among these possibilities, the singlet-resonance scenario requires
additional ingredients beyond those demonstrated by the benchmark
analysis. Its viability must therefore be assessed not only through
the predicted neutrino yield, but also through the consistency of
the required scalar spectrum, CP violation, and decay properties with 
theoretical and experimental checks. For example, the singlet--doublet mixing must be large
enough for $h_s$ to decay within the solar volume, yet small enough to evade the
constraints from rare $B$-meson decays, the BaBar limit on
$\Upsilon \to \gamma h_s$, and the exotic decay modes of the $125~\mathrm{GeV}$
Higgs boson. The CP violation in the singlet sector must be reconciled with
electric-dipole-moment limits, which appears feasible if the phase resides in a
nearly decoupled singlet sector. Finally, the parametric estimate of
Eq.~(\ref{eq:Nnu_scaling}) should be confirmed by a dedicated calculation
that includes capture, thermalization, the annihilation branching fractions, and
neutrino propagation in the Sun.

With these caveats in mind, the essential conclusion of this section is that
the GNMSSM supplies qualitatively new ingredients, absent in the MSSM, for
reconciling the LZ excess with the solar neutrino bounds: a reduced capture
rate at lower DM masses, a possibly inefficient thermalization history, and the
option of neutrino-poor annihilation final states. Whether these modifications,
individually or in combination, suffice to reconcile the LZ-favored parameter
region with the IceCube and Super-Kamiokande limits remains an open
quantitative question, which motivates dedicated investigations in future work.

\section{Conclusion}
\label{sec:conclusion}

The single nuclear recoil event at $E_R\simeq 248~\mathrm{keV}$ reported by the
LZ collaboration, if interpreted as a DM signal, calls for an endothermic
scattering process with a sub-MeV mass splitting between two nearly degenerate
DM states. Higgsino DM offers one of the few well-motivated realizations of
this picture, since the $Z$-mediated transition
$\widetilde\chi^0_1 N\to\widetilde\chi^0_2 N$ has a fixed, gauge-strength cross section
and the small splitting $\delta\simeq 350~\mathrm{keV}$ emerges from the high-energy
scale broken of an $U(1)$ Higgsino-number symmetry. Within the MSSM, however, 
this interpretation is
subject to three rigid correlations: the thermal relic abundance fixes the
Higgsino mass at about $1.1~\mathrm{TeV}$; the splitting is controlled solely
by the Gaugino masses, so that $\delta\sim\mathcal{O}(100)~\mathrm{keV}$
requires either $|M_{1,2}|\sim 10^{6}$--$10^{7}~\mathrm{GeV}$ or a Bino--Wino
cancellation tuned to per-mille accuracy at a few TeV; and the inelastic cross
section is fixed at its pure-Higgsino value. The last two properties, combined
with the fixed mass, render the scenario particularly vulnerable to the IceCube
and Super-Kamiokande limits on neutrinos from DM captured in the Sun, which
exclude $\delta\lesssim 566~\mathrm{keV}$ for a thermal $1.08~\mathrm{TeV}$
Higgsino under standard assumptions. The scientific question addressed in this
work is therefore whether these obstacles are intrinsic to the Higgsino
hypothesis or are artefacts of its minimal realization. Answering this question
is a prerequisite for deciding whether Higgsino DM remains a viable
interpretation of the LZ event, and for identifying which extensions of the
MSSM are worth pursuing should the event be confirmed.

The physical idea pursued here is that an independent source of
Higgsino-number breaking can decouple the three correlations above. The GNMSSM
supplies such a source through the singlet superfield $\widehat S$: the coupling
$\lambda\widehat S\widehat H_u\cdot\widehat H_d$, together with the general singlet mass
parameters, induces a Singlino-mediated contribution to the neutral-Higgsino
splitting whose magnitude and sign are governed by $\lambda$ and $d$ rather
than by $M_1$ and $M_2$. This contribution interferes with the Gaugino-induced
term, so that the condition $\delta\simeq 350~\mathrm{keV}$ becomes a matching
condition between the Gaugino and singlet sectors instead of a constraint on
the Gaugino masses alone. The same Higgsino--Singlino mixing rescales the
inelastic cross section by $\cos^2\theta$ and, together with Slepton
coannihilation, modifies the freeze-out dynamics. To examine these ideas
concretely, we analyzed the neutralino sector of the GNMSSM analytically,
including the parametric origin of the small splitting and its sensitivity to
the input parameters, and we confronted six benchmark points from our previous
global study, BP1--BP6, with the LZ event through an extended unbinned
likelihood in the $(|\mu_{\rm tot}|,\delta)$ plane. All six points satisfy the
constraints from DM direct detection experiments, Higgs precision data,
flavor observables $Br(B_s\to\mu^+\mu^-)$ and $Br(B \to X_s \gamma)$, Fermi-LAT dwarf
galaxies, and LHC searches for compressed electroweakinos and Sleptons.

The numerical results confirm the expected relaxation of the MSSM
correlations. Each benchmark point realizes $\delta\simeq 330$--$350~\mathrm{keV}$
with Gaugino mass parameters of $2$--$5~\mathrm{TeV}$, for both $M_1M_2>0$
and $M_1M_2<0$; the DM mass spans
$m_{\widetilde\chi^0_1}\simeq 600$--$1100~\mathrm{GeV}$, extending well below the
MSSM value while reproducing the Planck relic density; the Singlino fraction
$|N_{15}|^2$ ranges from nearly zero to the ten-percent level, reducing the
inelastic cross section to $\sigma_n^\text{GNMSSM}\simeq 6$--$7.2\times 10^{-39}~\mathrm{cm}^2$
and suppressing the Higgs-mediated elastic rate relative to the pure-Higgsino
limit; and all points lie within the $1\sigma$ region of the LZ likelihood,
with $\Delta\chi^2<1$ throughout and $\Delta\chi^2=0.03$ for BP6. We stress
that this flexibility does not remove the fine tuning associated with the
sub-MeV splitting: the sensitivity measures of all benchmark points exceed
$5\times 10^{3}$, reflecting a cancellation among contributions that are
individually several orders of magnitude larger than $\delta$. What the
GNMSSM achieves is a relocation of this tuning from the Gaugino sector, where
it would dictate an intermediate scale or a special relation between $M_1$ and
$M_2$, to a matching condition involving parameters that are otherwise free.
Since $\delta/m_{\widetilde\chi^0_1}\lesssim 10^{-6}$ and no symmetry is known to
enforce such a hierarchy, a comparable degree of tuning appears to be a generic
feature of inelastic-DM interpretations of the LZ event rather than a
peculiarity of the present framework.

The solar neutrino constraints, which constitute the most serious challenge to
the minimal Higgsino interpretation, have not been imposed on the benchmark
points, because the published limits rely on assumptions that need not hold in
the present scenario. We identified three mechanisms, specific to the GNMSSM,
through which they may be relaxed. First, the lower DM masses and reduced
$\cos^2\theta$ factor of BP1--BP6 tend to decrease the capture rate through both the
less efficient endothermic scattering and the kinematic closure of the inelastic channel over
a larger fraction of the solar volume. Second, once the inelastic channel has
closed, thermalization of the captured population proceeds through elastic
scattering, which is strongly suppressed in some SUSY realizations of the LZ
event; inefficient thermalization leaves the captured DM spatially diffuse,
reduces the annihilation rate, and may prevent capture--annihilation
equilibrium. Third, in a distinct region of parameter space with
$m_{\widetilde\chi^0_1}\gg 1.1~\mathrm{TeV}$, resonant annihilation through a
CP-violating singlet pseudoscalar into light singlet scalars can render the
final states neutrino-poor. Whether these mechanisms, individually or in
combination, reconcile the LZ-favored region with the IceCube and
Super-Kamiokande limits is an open quantitative question that requires a
dedicated treatment of capture, non-equilibrium thermalization, and neutrino
propagation in the Sun.

In summary, the obstacles that confront a Higgsino interpretation of the LZ
high-recoil event in the MSSM originate in the rigid correlations of the
minimal model rather than in the Higgsino hypothesis itself. A singlet
extension with an independent source of Higgsino-number breaking accommodates
the event with multi-TeV Gauginos of either relative sign and with Higgsino
masses below $1~\mathrm{TeV}$, while satisfying all terrestrial constraints
considered. The resulting scenario is testable: the predicted inelastic rate
lies within reach of the ongoing XENONnT and  PandaX-4T exposures, the compressed
Higgsino spectrum with $\mathcal{O}(100)~\mathrm{keV}$ splitting and nearby
Sleptons is a target for the High-Luminosity LHC and dedicated long-lived
particle searches, and the solar neutrino signal, once evaluated with the
nonconventional dynamics described above, provides a complementary probe
through next-generation neutrino telescopes.

\appendix

\section*{Acknowledgement}

This study emerged from extensive and rigorous collaborative
 discussions among all three authors, who are listed 
 alphabetically by surname. F. Li conducted the majority 
 of the computational analyses. J. Cao conceived the research 
 framework and carried out the parameter-space scans. S. Bisal 
 constructed the likelihood function for the LZ high-recoil 
 event analysis. The manuscript was jointly drafted and 
 revised by all authors.

We thank Yang Zhang for helpful discussions. This work was supported by 
the National Natural Science Foundation of China (NSFC) under Grant No.~12575110.


\bibliographystyle{CitationStyle}
\bibliography{myrefs}
\end{document}